\documentclass{PoS}
\usepackage{wrapfig}   
\usepackage[percent]{overpic} 
\usepackage{tikz}

\title{Hot strong matter}

\ShortTitle{Hot strong matter}

\author{\speaker{Katarzyna Grebieszkow} 
\\
        Warsaw University of Technology\\
        E-mail: \email{kperl@if.pw.edu.pl}}

\abstract{The recent results on relativistic heavy-ion collisions are discussed. The most convincing quark-gluon plasma signatures at the LHC and the top RHIC energies are presented. Moreover, the possible methods of evaluating the energy threshold for deconfinement (onset of deconfinement) are described, and the corresponding results from the RHIC Beam Energy Scan and the SPS programs are shown. Additionally, the first signatures of creating dense and collectively behaving systems in collisions of small nuclei (or even in elementary interactions) are discussed. Finally, the current status of experimental search for the critical point of strongly interacting matter is summarized.}     

\FullConference{XXII. International Workshop on Deep-Inelastic Scattering and Related Subjects,\\
		28 April - 2 May 2014\\
		Warsaw, Poland}

\begin{document}

\section{Introduction}

It is a well established fact that matter exists in different states. For strongly interacting matter at least three are expected: normal nuclear matter (liquid), hadron gas (HG), and a system of deconfined quarks and gluons (eventually the quark-gluon plasma, QGP). In cosmology, it is believed that the early Universe consisted of QGP $\sim$few microseconds after the Big
Bang. Nowadays, QGP may exist in the core of neutron stars. One of the most important goals of high-energy heavy-ion collisions is to establish the phase diagram of strongly interacting matter by finding the possible phase boundaries and critical points. In principle, we want to produce the quark-gluon plasma and analyze its properties and the transition between QGP and HG. This goal seems to be truly attractive nowadays, when three heavy-ion accelerators are working and taking data. These are: Super Proton Synchrotron (SPS) at CERN, Relativistic Heavy Ion Collider (RHIC) at BNL and Large Hadron Collider (LHC) at CERN. A very hot and high energy density system is created at the top RHIC ($\sqrt{s_{NN}}$=200 GeV) and current LHC ($\sqrt{s_{NN}}$=2.76 TeV) energies. Therefore, these accelerators are best suited to investigate the properties of quark-gluon plasma. The region of the transition between hadron gas and QGP has been studied by the SPS experiments since more than twenty years, and from 2010 it has been also studied within the Beam Energy Scan (BES) program of the RHIC accelerator.

\begin{wrapfigure}{r}{6.cm}
\centering
\vspace{-0.5cm}
\includegraphics[width=0.4\textwidth]{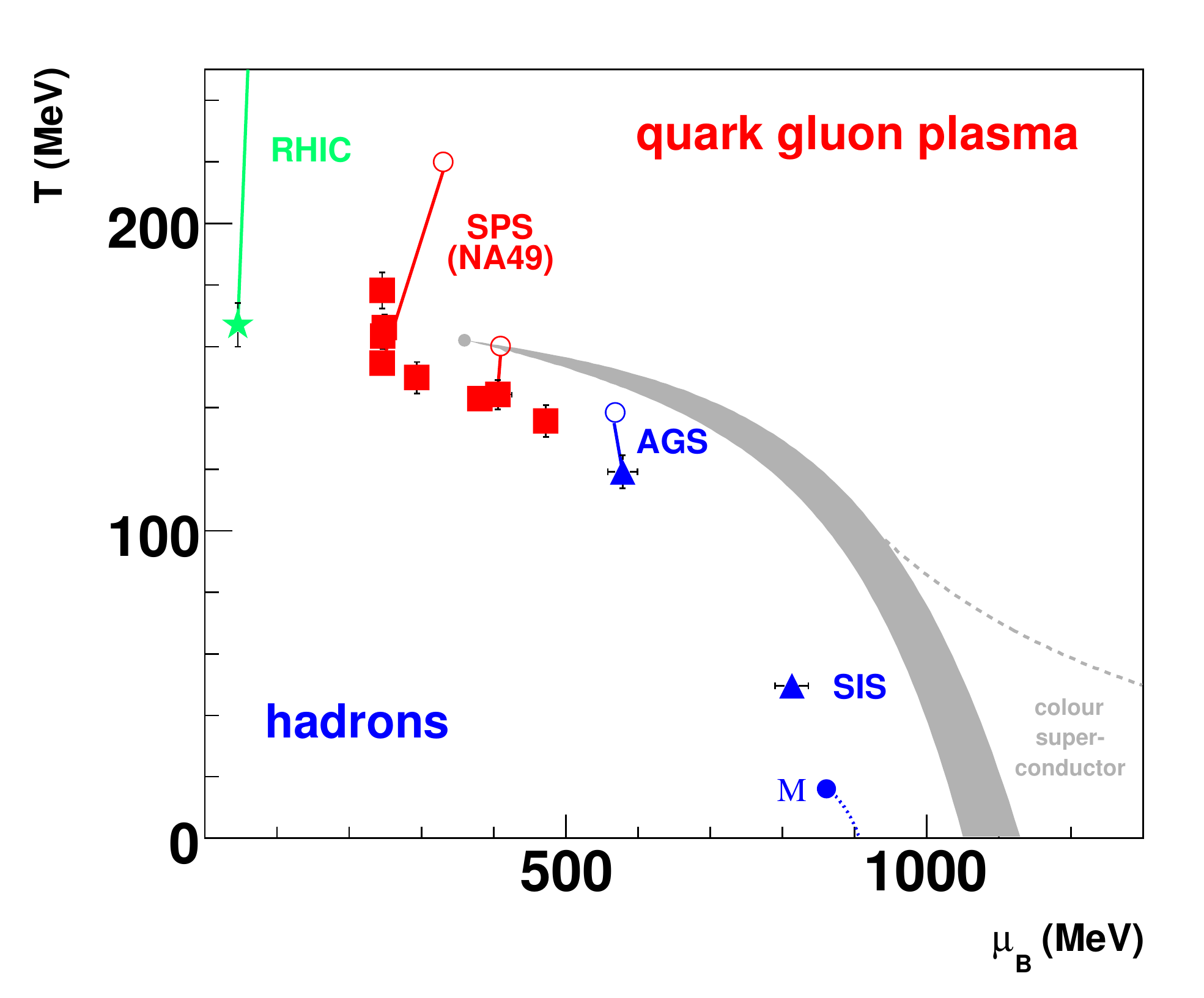}
\vspace{-1cm}
\caption[]{\footnotesize {Phase diagram of strongly interacting matter.}}
\label{phas_full}
\end{wrapfigure}

The phase diagram of strongly interacting matter (Fig.~\ref{phas_full}) is most often presented in terms of temperature ($T$) and baryochemical potential ($\mu_B$), which reflects net-baryon density. It is believed that for large values of $\mu_B$ the phase transition is of the first order (gray band) and for low $\mu_B$ values a rapid but continuous transition (cross-over). A critical point of second order (CP) separates those two regions. The open points in Fig.~\ref{phas_full} are hypothetical locations reached by the high-density matter droplet after dissipation of the energy of the incident nucleons from where the evolution of the expanding and cooling fireball starts. The temperatures in these points can reach from 230 MeV to about 600 MeV (from the top SPS to current LHC energies; $T=173$ MeV $\approx 2 \cdot 10^{12}$ K) and the energy densities from about 3 GeV/fm$^3$ at the top SPS energy to more than 5 and 15 GeV/fm$^3$ at the top RHIC and LHC energies\footnote{These Bjorken's estimates of energy densities at RHIC and LHC should be treated as lower limits (hydrodynamical models give here much higher values, especially at LHC energies). Moreover, in these estimates the so-called thermalization time was assumed to be $\tau_{0}=1$ fm/c for all energies (Bjorken's energy density is proportional to $1/\tau_{0}$), whereas it is rather commonly believed that at higher energies $\tau_{0}$ should be smaller (most probably $\sim$0.6 fm/c at the top RHIC energy).}, respectively (the energy density of normal nuclear matter is about 0.16 GeV/fm$^3$).     
The closed symbols in Fig.~\ref{phas_full} represent chemical freeze-out points \cite{beccatini} (chemical composition fixed; the end of inelastic interactions). The fireball then expands further until thermal (kinetic) freeze-out takes place (particle momenta are fixed). As the temperature (and energy density) decreases during the expansion, the chemical freeze-out temperature ($T_{chem}$) is higher than the kinetic ($T_{kin} \equiv T_{fo}$) one.

The energies of the CERN SPS cover a very important region of the phase diagram. First, the NA49 experiment showed that the energy threshold for deconfinement (minimum energy to create a partonic system) is located at low SPS energies ($\sqrt{s_{NN}} \approx 7.6$~GeV or beam energy 30$A$ GeV; see the open point hitting the 
transition line in Fig.~\ref{phas_full}) \cite{na49pikp}. Second, theoretical calculations suggest that the critical point of strongly interacting matter may be located at energies accessible at the CERN SPS. There are several such predictions available nowadays, let us mention the examples: 
$(T^{CP}, \mu_B^{CP}) = (162(2), 360(40))$ MeV \cite{fodor_latt_2004} 
($\mu_B=360$ MeV corresponds to $\sqrt{s_{NN}}=9.7$ MeV \cite{beccatini}), 
or $(T^{CP}, \mu_B^{CP}) = (0.927(5)T_c, 2.60(8)T_c) = (\sim 157, \sim 441)$ MeV \cite{lat_2011}, 
or $(T^{CP}/T_c, \mu_B^{CP}/T^{CP}) = (\sim 0.96, \sim 1.8)$ ($\mu_B \sim 290$ MeV) \cite{latt_qm12}, 
where $T_c$ is the cross-over temperature at $\mu_B=0$. 
It should be, however, stressed that lattice calculations ruling out the CP existence, and showing cross-over transition in the whole $\mu_B$ range, are also available on the market \cite{no_cp}. Therefore, the experimental results may be crucial in the verification of existing theoretical predictions.

\section{LHC and top RHIC energies: ''QGP desert''}

\begin{figure}[h]
\centering
\includegraphics[width=0.41\textwidth]{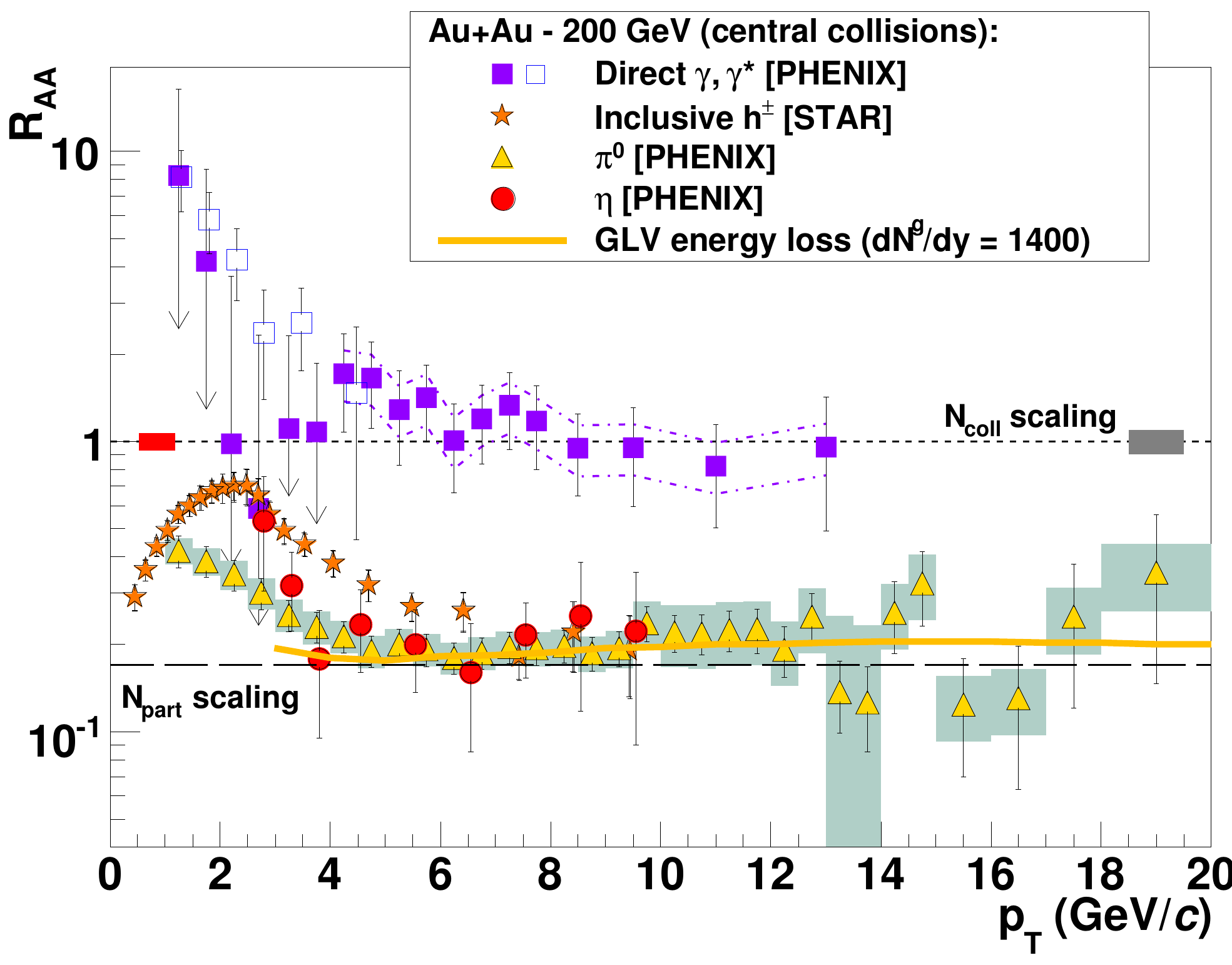}
\begin{overpic}[width=0.47\textwidth]{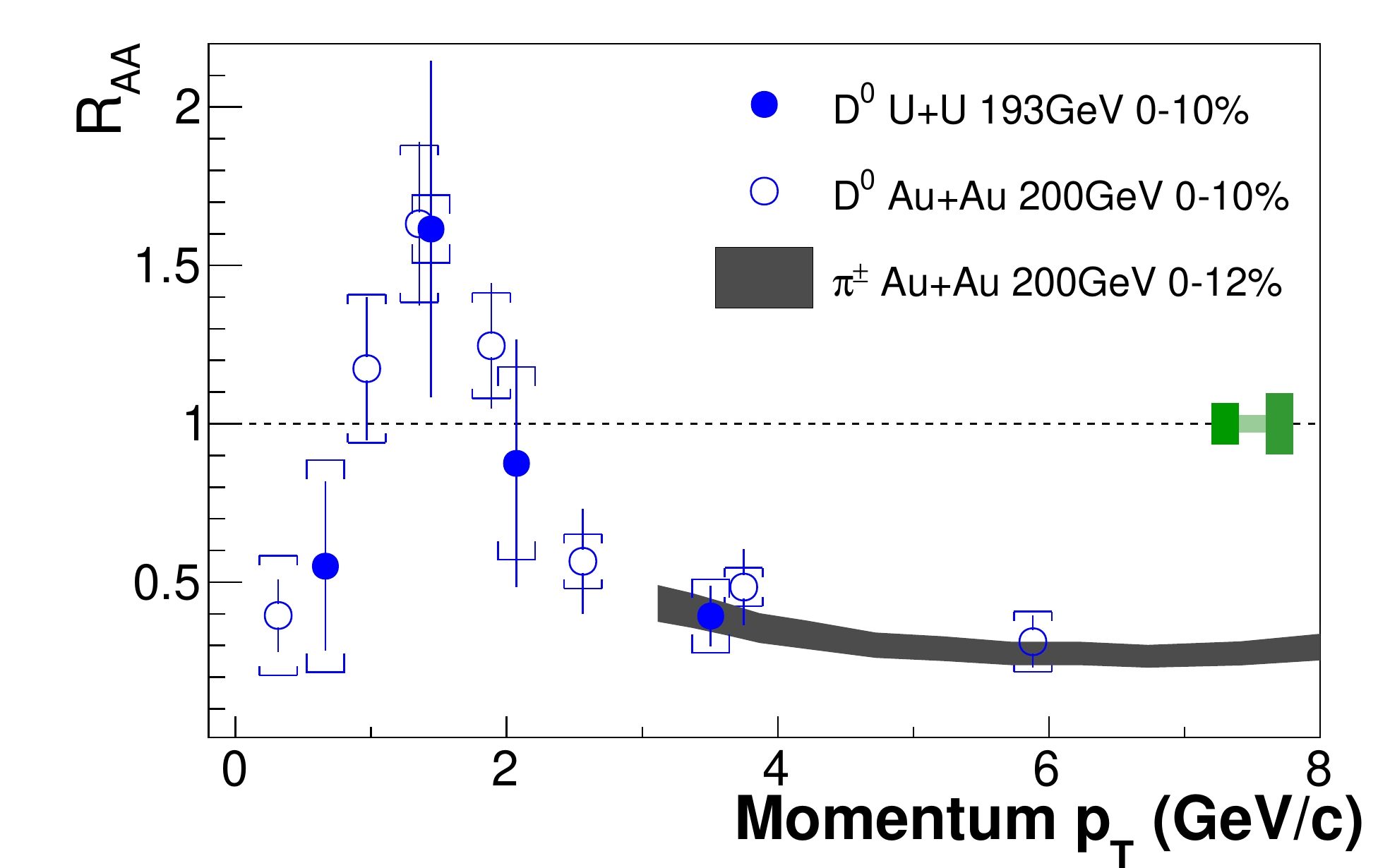}
        \put(15,60) {\footnotesize U+U data: $STAR$ $Preliminary$}
\end{overpic}
\vspace{-0.1cm}
\caption[]{\footnotesize {Left: $R_{AA}$ of different particle species in $Au+Au$ collisions at the top RHIC energy ($\sqrt{s_{NN}}=200$ GeV). Figure taken from \cite{Enterria}. Right: $R_{AA}$ of $D^0$ mesons in $Au+Au$ \cite{STAR_D0_RAA} and $U+U$ \cite{trzeciak_WWND14} collisions at top RHIC energies. Figure prepared by Zhenyu Ye (STAR Collab.). }} 
\label{RAA_topRHIC}
\end{figure}

If we want to study QGP properties we should focus on LHC and top RHIC energies. The most famous QGP signals at the top RHIC energy are jet quenching and scaling of elliptic flow with the number of constituent quarks. Jet is a ''spray'' of collimated particles originating from fragmentation of hard-scattered quark or gluon. There are several ways of studying jets: using algorithms of full jet reconstruction (but they are much more effective in elementary interactions), two-particle correlations in azimuthal angle, and studying nuclear modification factor. The last two are {\it indirect} methods of studying jets but since recently these have been the only ways of jet analysis in case of nucleus+nucleus interactions. The nuclear modification factor ($R$) is defined as:
\begin{equation}
R_{AA}(p_T) = \frac {1} {N^{AA}_{coll}} \frac {(Invariant\; yield)_{AA}} {(Invariant\; yield)_{pp}},
\end{equation}
where $N^{AA}_{coll}$ is the number of binary ($N+N$) collisions obtained from the Glauber model. If we want to compare central and peripheral $A+A$ collisions we can use:
\begin{equation}
R_{CP}(p_T) = \frac {N^{PERIPH}_{coll}} {N^{CENTRAL}_{coll}} \frac {(Invariant\; yield)_{CENTRAL}} {(Invariant\; yield)_{PERIPH}}.
\end{equation}
For soft processes (low-$p_T$ region) $R_{AA}$ should be smaller than one, because in this case particle production scales with the number of participants ($N_{part}$) or with the number of wounded nucleons ($N_W$)\footnote{Typically $N_{part}$ denotes the number of nucleons (from both nuclei) participating in the collision, whereas $N_W$ is the number of participants suffering at least one {\it inelastic} interaction.}. For hard processes (high-$p_T$ region) we expect that particle production scales with the number of binary collisions ($N_{coll}$) and thus $R_{AA}=1$. Therefore, in the absence of nuclear effects we expected that $R_{AA}$ would increase and then saturate at the value of 1. In some interactions ($p+A$, $d+Au$, etc.) $R_{AA}$ was found to be higher than one at intermediate $p_T$ region. This so-called Cronin effect is interpreted as probably due to initial elastic multiple low-momentum scattering of the parton (from projectile nucleon) on target nucleons. Thus, before the final hard parton+parton interaction the ''projectile parton'' already has got its $p_T$ higher than zero. In central $Au+Au$ collisions at the top RHIC energy the long-awaited \cite{pt_sup_th} suppression of high-$p_T$ particles was observed\footnote{See Quark Matter 2014 conference slides for the most recent results on quenching of whole jets (full jet reconstruction) both at RHIC and LHC energies.}. Figure~\ref{RAA_topRHIC} (left) shows that this suppressions was a factor of five ($R_{AA}$ at high $p_T$ close to 0.2), but was not seen for photons (they do not interact strongly with the medium) thus confirming that the observed suppression is a final state effect, which was interpreted as due to parton energy loss while traveling through hot and dense medium produced in $A+A$ collision. The orange solid line in Fig.~\ref{RAA_topRHIC} (left) shows the theoretical predictions for parton energy loss in a medium with gluon density per rapidity unit $dN^g/dy \approx 1400$, which corresponds to the temperature of the medium $T \approx 400$ MeV \cite{Enterria}. The observed jet attenuation ({\it jet quenching}) is an evidence of the extreme energy loss of quarks or gluons traversing a large density of color charges and reflects the extreme opacity of the QGP.

\begin{figure}[h]
\centering
\includegraphics[width=0.27\textwidth]{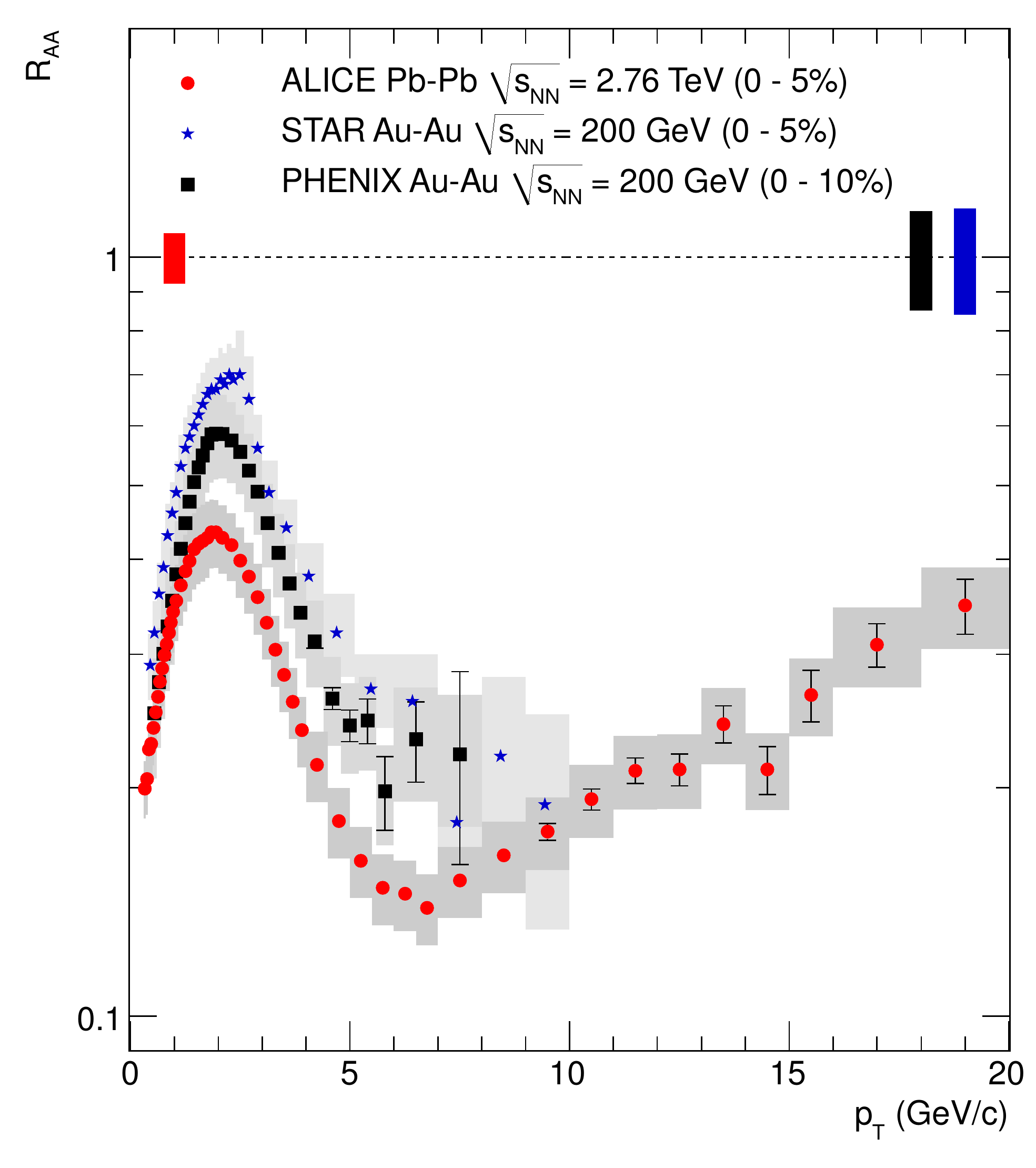}
\includegraphics[width=0.37\textwidth]{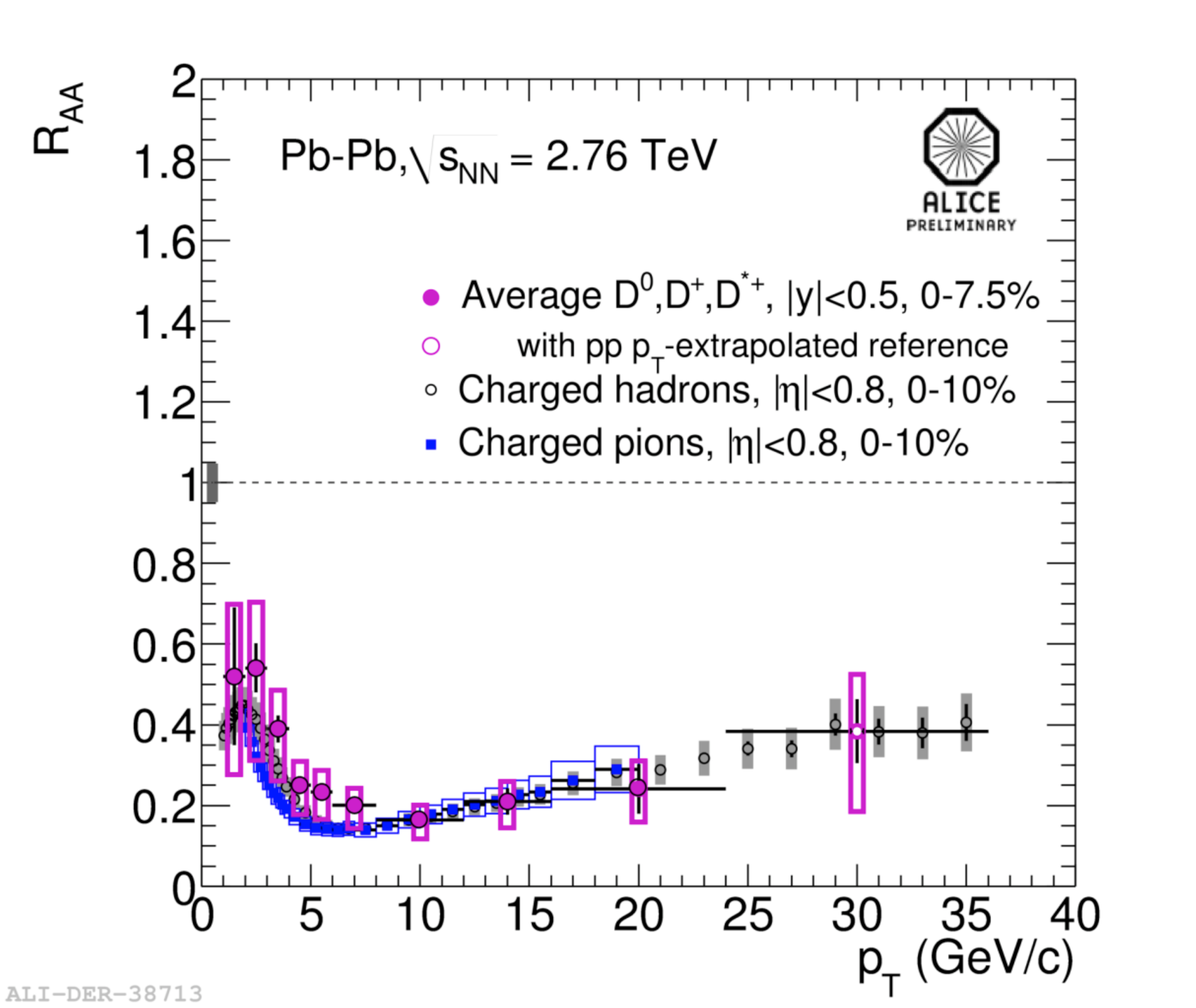}
\includegraphics[width=0.3\textwidth]{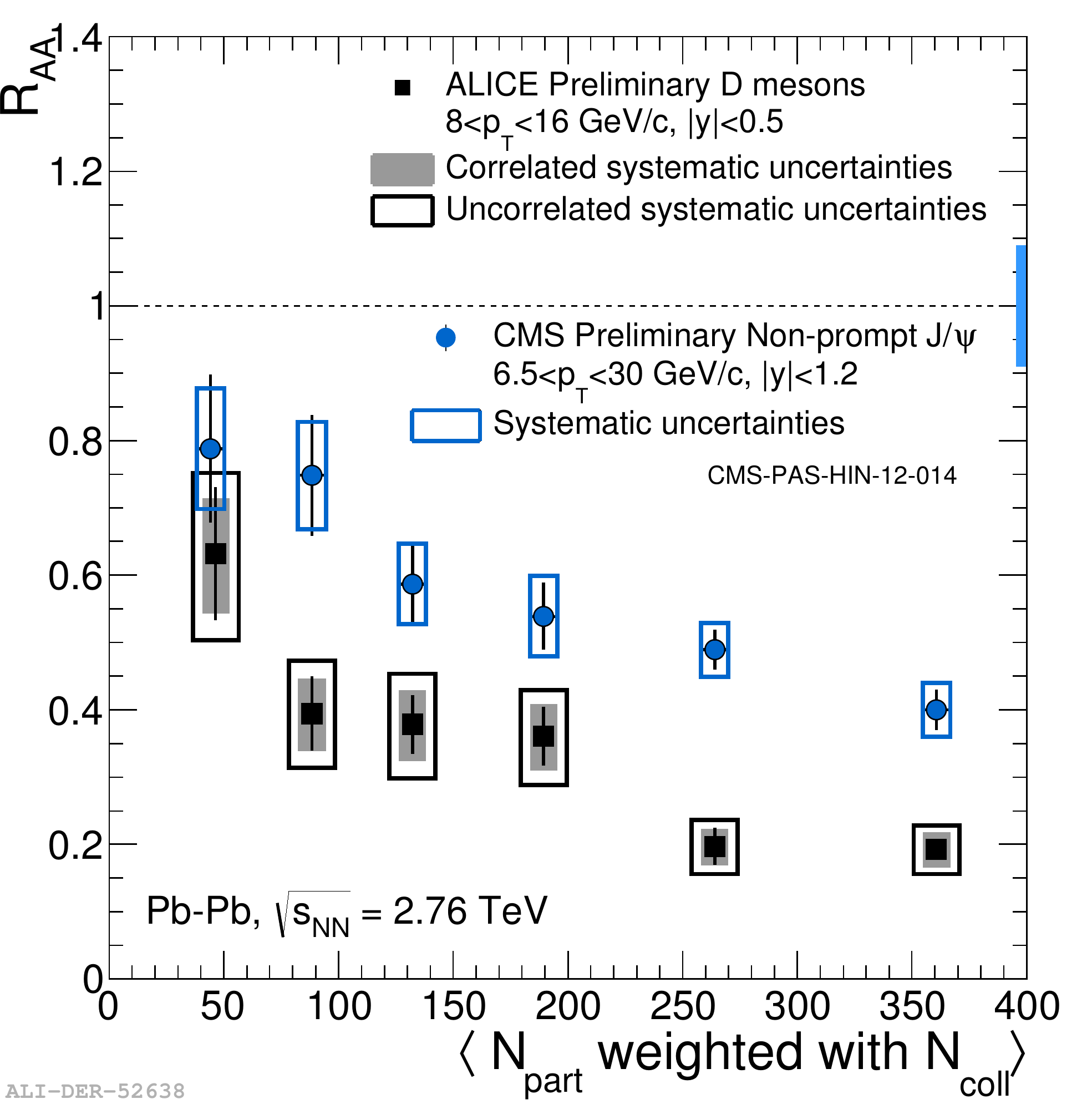}
\caption[]{\footnotesize {$R_{AA}$ of charged particles (left) \cite{alice_Raacharged}, $D$ mesons (middle) \cite{alice_RaaD}, and $J/\Psi$ particles coming from $B$ meson decays \cite{cms_RaaJpsi} versus transverse momentum or number of participants $N_{part}$ (measure of centrality; high $N_{part}$ represents most central collisions). Results are shown for $Pb+Pb$ collisions at LHC ($\sqrt{s_{NN}}=2.76$ TeV).}}
\label{RAA_LHC}
\end{figure}

The theory expected that the energy losses (for $E_{parton}=const.$) due to induced gluon radiation in a dense color medium should follow hierarchy: 
$\Delta E_{rad}(g) > \Delta E_{rad}(q_{light}) > \Delta E_{rad}(c) > \Delta E_{rad}(b)$ \cite{energyloss_th}, but at the top RHIC energy the suppression seems to be similar for light and heavy particles. This can be seen in Fig.~\ref{RAA_topRHIC} (right) for $D^0$ mesons, which at high-$p_T$ region show suppression similar to that of the light particles. 
Figure~\ref{RAA_LHC} presents the nuclear modification factors of different particle species in $Pb+Pb$ collisions at the LHC. The suppression of high-$p_T$ particles (at minimum $p_T$=6-7 GeV/c) is even stronger than at the top RHIC (left plot). Moreover, heavy $D$ mesons are suppressed on a similar level as charged particles (middle plot). However, in contrary to the top RHIC energy, the suppression of beauty is smaller, it is $R_{AA}(D) \sim R_{AA}(\pi) \leq R_{AA}(B \rightarrow J/\Psi)$. In the case of larger energy loss for charm than for beauty ($\Delta E(c) > \Delta E (b)$) we can expect $R_{AA} (D) < R_{AA} (B)$. Therefore, Fig.~\ref{RAA_LHC} (right) is an indication of larger energy loss for charm than for beauty.

\begin{wrapfigure}{r}{6.cm}
\centering
\includegraphics[width=0.32\textwidth]{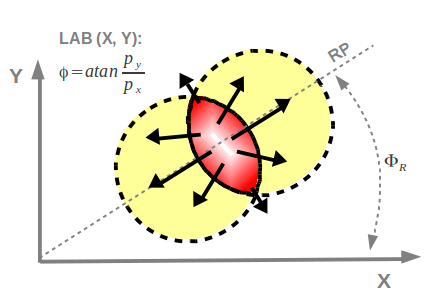}
\vspace{-0.3cm}
\caption[]{\footnotesize {Schematic view of non-central collision in a plane perpendicular to the beam axis (Z). The lengths of black arrows represent the strength of elliptic flow. }}
\label{flow_fig}
\end{wrapfigure}

Collectivity, and in principle the scaling of elliptic flow with the number of constituent quarks, is considered as one of the most important QGP signatures in $Au+Au$ collisions at the top RHIC energy. For non-central collisions the triple differential invariant distribution of particles emitted in the final state is Fourier-decomposed as follows:

\begin{equation}
E \frac {d^3 N} {dp^3} = \frac {1} {2 \pi} \frac {d^2 N} {p_T dp_T dy} ( 1 + 2 v_1 \cos (\phi - \Phi_R)+ 2 v_2 \cos [2(\phi - \Phi_R)] + ...),
\label{flow_eq}
\end{equation}
where $y$ is the rapidity, $\phi$ is the azimuthal angle of a particle in the laboratory frame (see Fig. \ref{flow_fig}), $\Phi_R$ is the azimuthal angle of the reaction plane (RP) in the laboratory system (different for each event), and $v_n = \langle \cos [n(\phi-\Phi_R)] \rangle$ are the Fourier coefficients. For central collisions we have radial flow only (the transverse shape is a circle instead of the almond-shaped overlap zone seen by a red color in Fig. \ref{flow_fig}). For non-central collisions radial flow is modulated by $v_n$ values (flow is anisotropic). In Equation~\ref{flow_eq} $1$ represents radial flow (isotropic emission in azimuthal angle; reaction plane cannot be defined), $v_1$ is the directed flow and $v_2$ the elliptic flow (higher harmonics can be neglected\footnote{In Equation \ref{flow_eq} the smooth almond-shaped overlap zone is assumed. See \cite{flow_intro} for the idea of recent calculations including initial state event-by-event fluctuations in participant positions.}). 
The idea of elliptic flow is schematically shown in Fig.~\ref{flow_fig}, where the initial spatial anisotropy (almond-shaped zone) via rescattering(!) is transformed into pressure gradients (higher in the reaction plane) which, in turn, are converted into anisotropy in momentum space (more particles are emitted "in-plane" than "out-of-plane" - see the black arrows in Fig. \ref{flow_fig}). The elliptic flow is sensitive to the early ($\sim$ fm/c) history of the collision. Higher pressure gradients "in-plane" imply that the expansion of the source would gradually diminish its anisotropy, making the $v_2$ so-called self-quenching variable. The elliptic flow builds up early, when the anisotropy is significant, and tends to saturate as the anisotropy continues to decrease. This is unlike the radial flow which continues to grow until freeze-out and is sensitive to both early and later times of the history of the collision. Therefore, $v_2$ is a signature of pressure at early times.

\begin{figure}[h]
\centering
\vspace{0.2cm}
\includegraphics[width=0.55\textwidth]{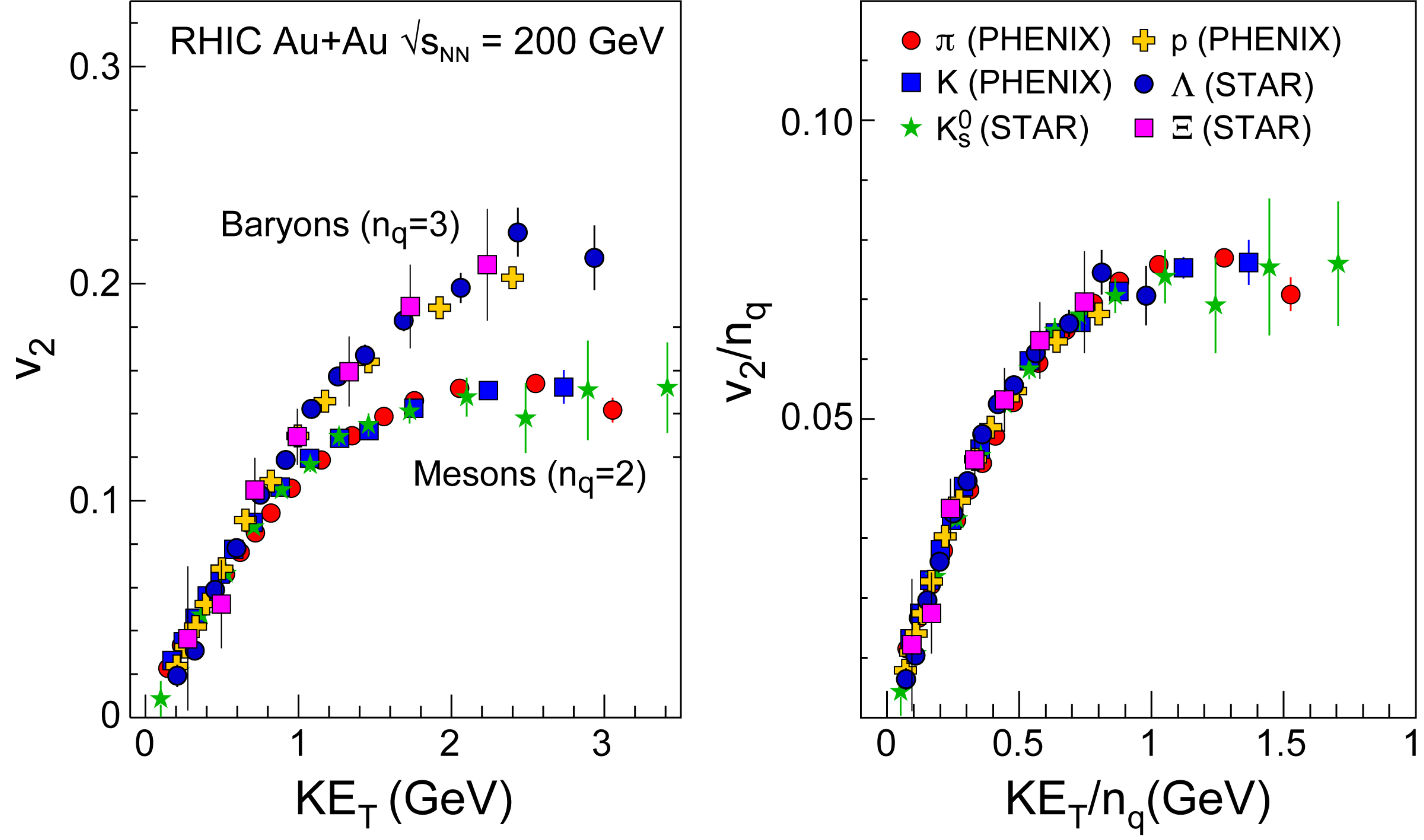}
\vspace{-0.2cm}
\caption[]{\footnotesize {Elliptic flow of baryons and mesons versus transverse kinetic energy $KE_T$ for $Au+Au$ collisions at $\sqrt{s_{NN}}=200$ GeV. In the right panel the two axes are divided by $n_q=3$ (baryons) or $n_q=2$ (mesons). Figure taken from \cite{Heinz_2008}.}}
\label{v2_scalingRHIC}
\end{figure}

In coalescence models of hadronization (they assume QGP phase!) elliptic flow of hadrons is inherited from elliptic flow of their constituent quarks (hadron flow behaves like a sum of flows of constituent quarks). It was proposed that: $v_2^{MESON}(p_T) \approx 2v_2^q(p_T/2)$ and $v_2^{BARION}(p_T) \approx 3v_2^q(p_T/3)$, where $v_2^q$ is the elliptic flow of a quark. Such a scaling with the number of constituent quarks $n_q$ (so-called NCQ scaling) was indeed observed in $Au+Au$ collisions at the top RHIC energy; it is even better seen in transverse kinetic energy $KE_T = m_T - m = \sqrt{m^2+p_T^2}-m$, where $m$ is the particle mass. In the left panel of Fig.~\ref{v2_scalingRHIC} two distinct branches are seen, one for baryons and the other for mesons. When the two axes are scaled by $n_q$ (right panel of Fig.~\ref{v2_scalingRHIC}), the two curves merge into one universal curve, suggesting that the flow is developed at the quark level, and hadrons form by the merging of constituent quarks. The NCQ scaling of $v_2$ is treated as a proof of the partonic collectivity (flow is originally developed on the quark level; quarks flow in QGP). The partonic collectivity is the key QGP signature at the top RHIC energies.

\begin{figure}[h]
\centering
\vspace{-0.6cm}
\includegraphics[width=0.6\textwidth]{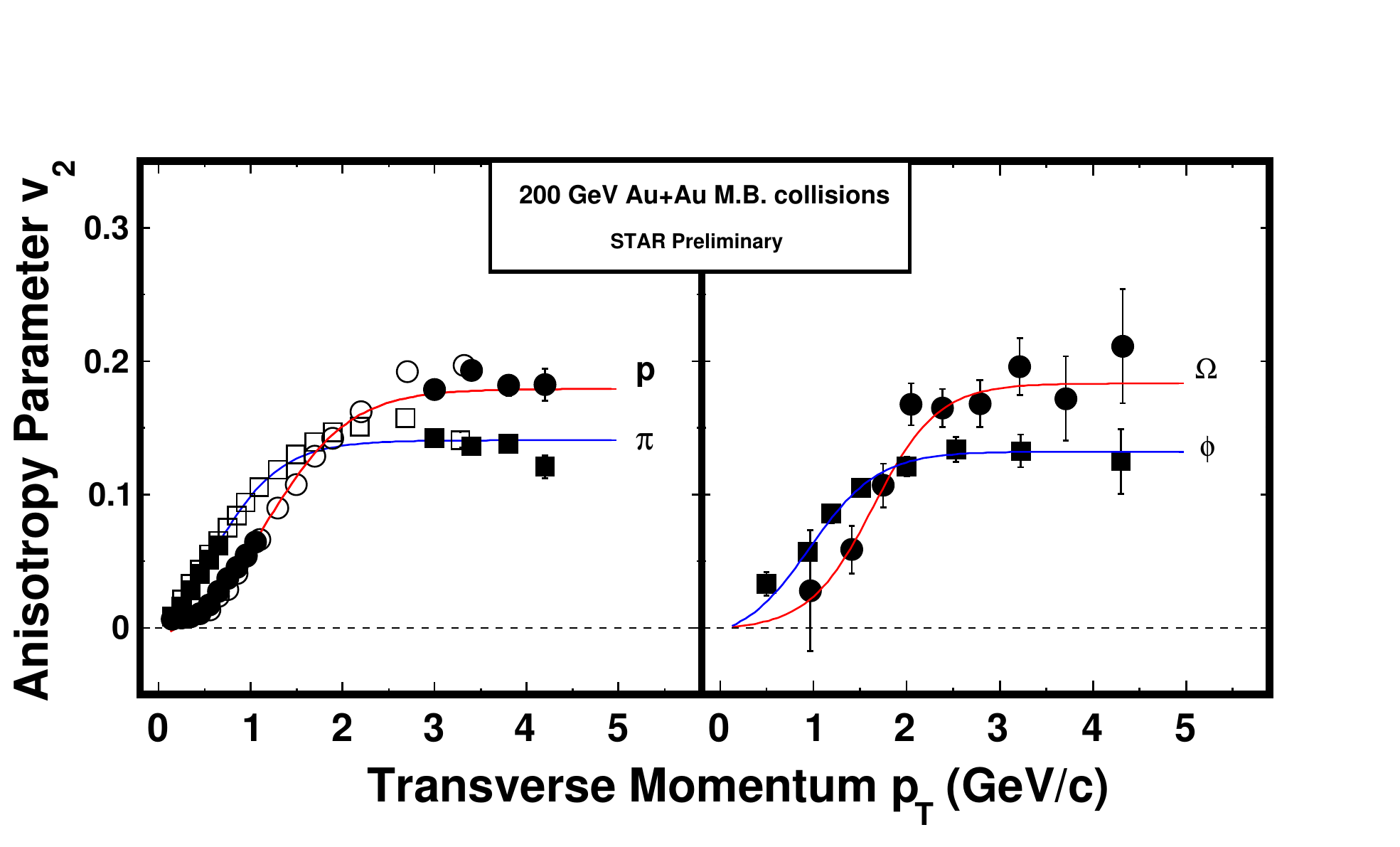}
\vspace{-0.4cm}
\caption[]{\footnotesize {Elliptic flow of different particle species measured in $Au+Au$ collisions at $\sqrt{s_{NN}}=200$ GeV \cite{v2_phi_omega_star}.}}
\label{v2_phi_omega}
\end{figure}

The another evidence of the partonic collectivity can be seen in Fig.~\ref{v2_phi_omega}, where $v_2$ of heavier (including 's' quarks) particles is compared with $v_2$ of pions and protons. The $\phi$ mesons are as heavy as protons but their $v_2$ is similar to $v_2$ of $\pi$ mesons, which suggests that $n_q$ is important! Not a particle mass. The fact that 's' quarks flow similarly to light 'u' and 'd' quarks is an another argument for the partonic collectivity (developed mostly in a deconfined phase). Moreover, $\phi$ and $\Omega$ particles have small cross sections for hadronic interactions and probably freeze-out earlier, thus they are promising observables of the early stage (they should be less affected by later stage hadronic interactions). Therefore, the results on $v_2$ suggest that the significant part of collectivity was developed in a partonic stage.

\begin{figure}[h]
\begin{tikzpicture}
    \begin{scope} [xshift=-2cm]
    \node {\includegraphics[width=0.325\textwidth]{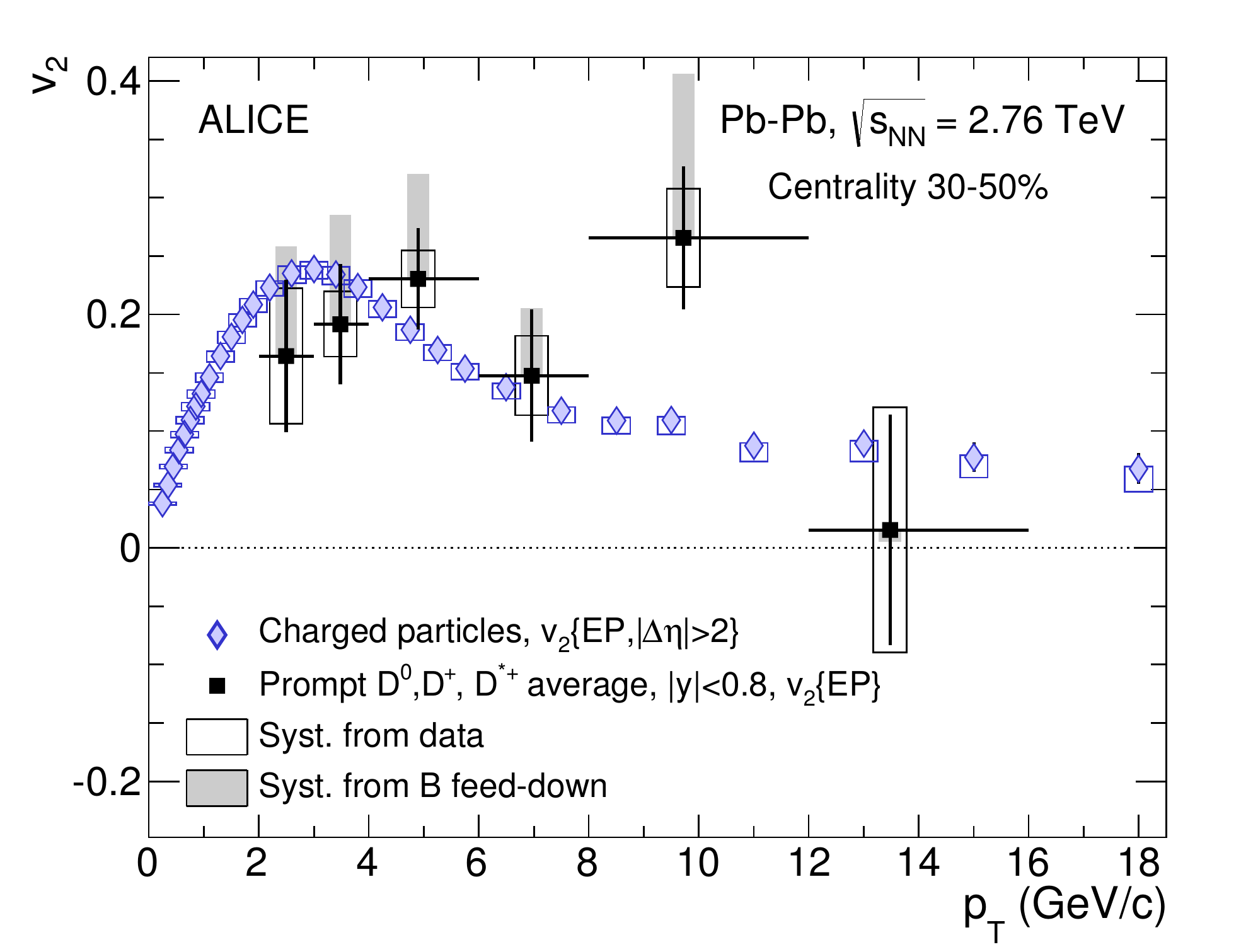}};
    \end{scope}
    \begin{scope} [xshift=3cm, yshift=-0.13cm]
    \node {\includegraphics[width=0.325\textwidth]{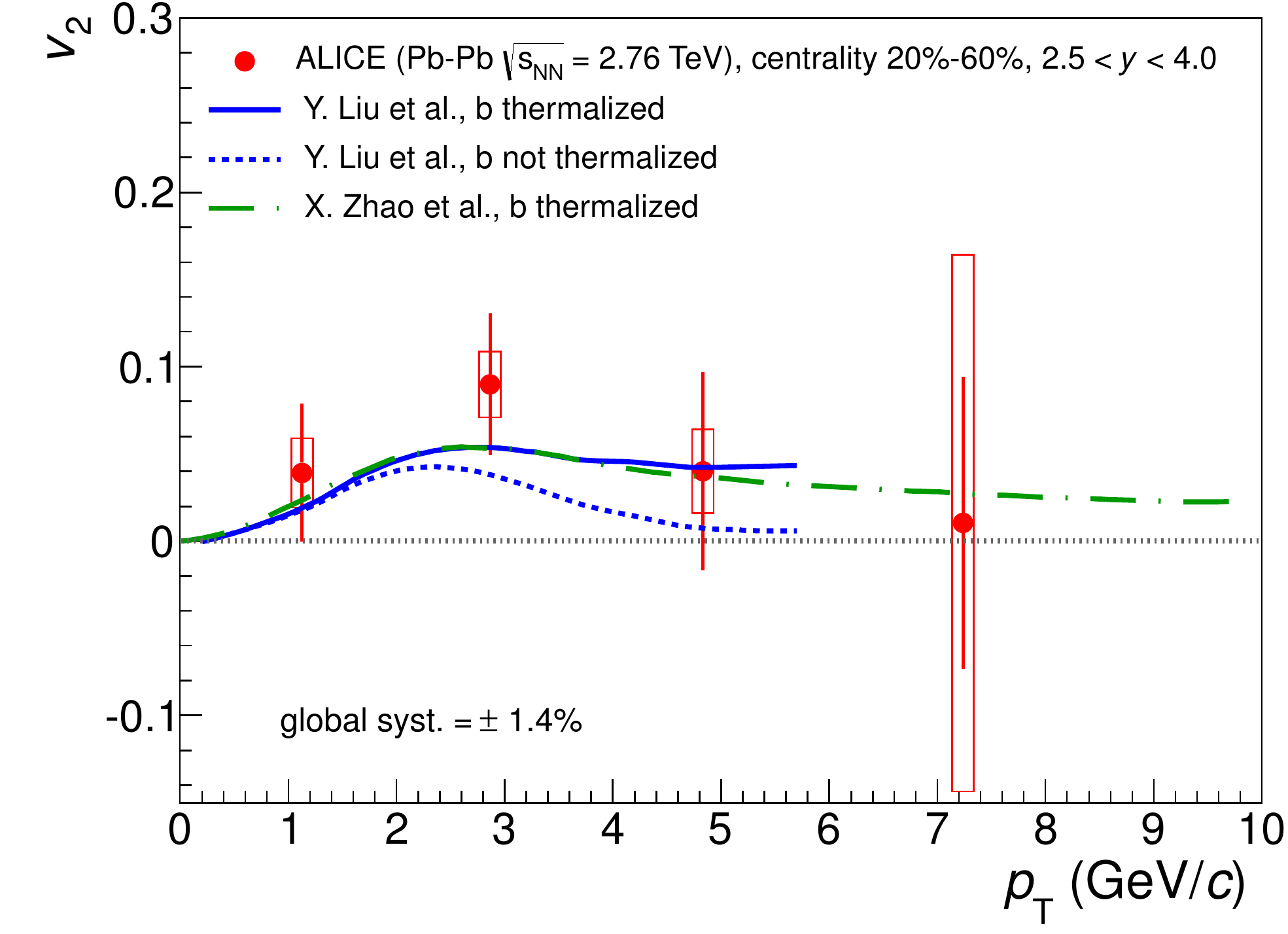}};
    \end{scope}
    \begin{scope} [xshift=8cm,  yshift=+0.1cm] 
    \node {\includegraphics[width=0.33\textwidth]{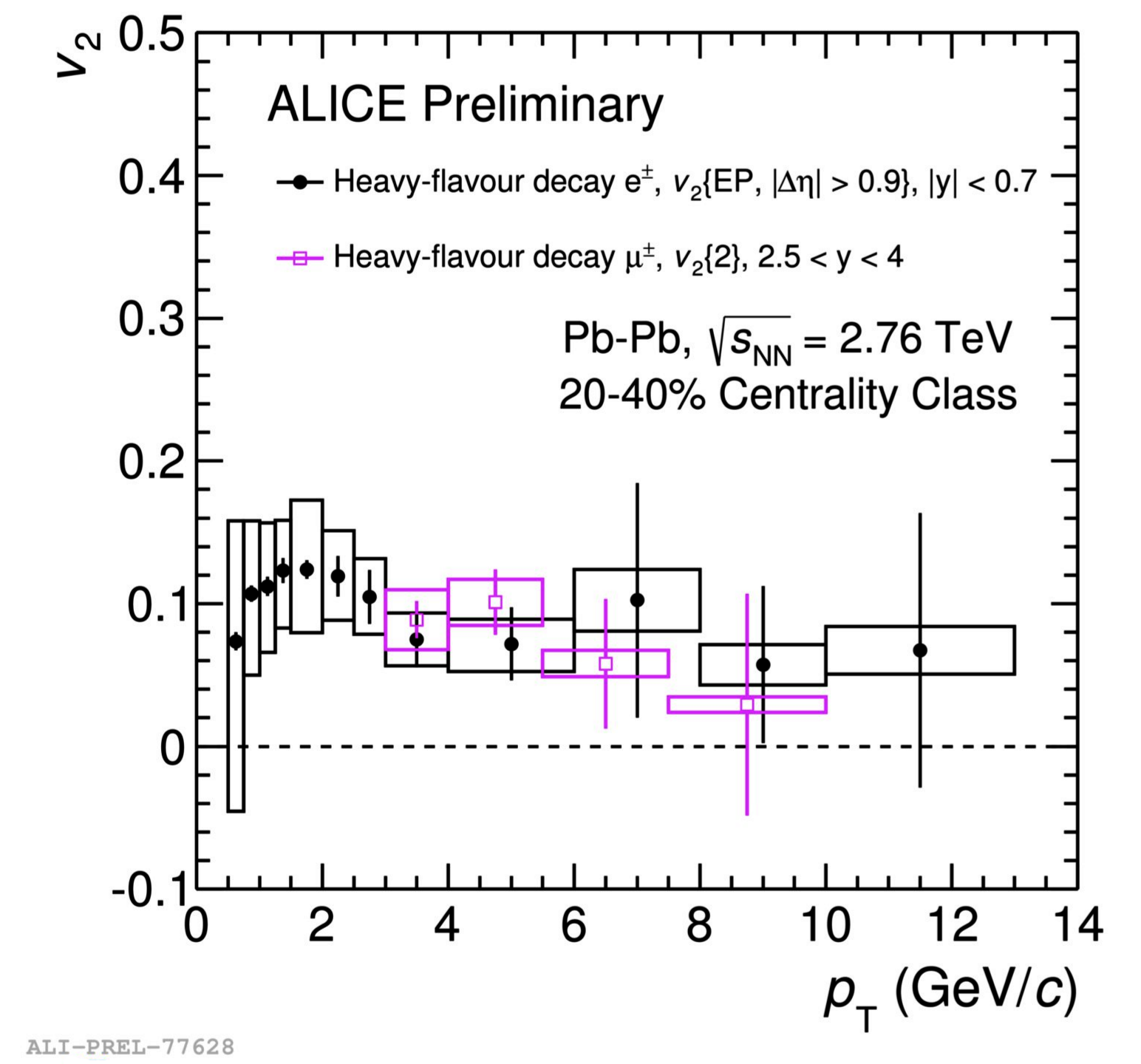}};
    \end{scope} 
\end{tikzpicture}
\vspace{-1cm}
\caption[]{\footnotesize {Elliptic flow of charged particles and $D$ mesons (left) \cite{v2_Dmesons_ALICE}, $J/\Psi$ mesons (middle) \cite{v2_jpsi_alice} 
, and leptons from heavy flavour decays (right) \cite{RBailhache} measured in $Pb+Pb$ collisions at $\sqrt{s_{NN}}=2.76$ TeV. }}
\label{v2_HF_LHC}
\end{figure}

Figure~\ref{v2_HF_LHC} presents the elliptic flow of heavy flavours: $D$ mesons (left), $J/\Psi$ mesons (middle), and leptons from heavy flavour decays (right) measured in $Pb+Pb$ collisions at the LHC energy. Elliptic flow of $D$ mesons is similar to that of the light particles suggesting that charm quarks participate in the collective flow of the expanding medium. Moreover, the $v_2$ of hidden charm ($J/\Psi$ mesons) is positive (in $Au+Au$ collisions at the top RHIC energy it was consistent with zero \cite{v2_jpsi_star}). Finally, Fig.~\ref{v2_HF_LHC} (right) shows that heavy-flavour decay muons and electrons (from 'c' and 'b' decays) also experience the anisotropic expansion of the medium ($v_2> 0$); the same observation was done for heavy-flavour decay electrons at the top RHIC energy \cite{v2_hf_star}. Due to their huge masses ($m_c \sim 1.3$ GeV, $m_b \sim 5$ GeV) the thermalization and flow of heavy quarks was much less probable. The results shown in Fig.~\ref{v2_HF_LHC} suggest that heavy quarks may be thermalized and may participate in the collective motion of the system! It is sometimes compared to the stones flowing with the stream.

\vspace{0.5cm}

\begin{figure}[h]
\centering
\includegraphics[width=0.34\textwidth]{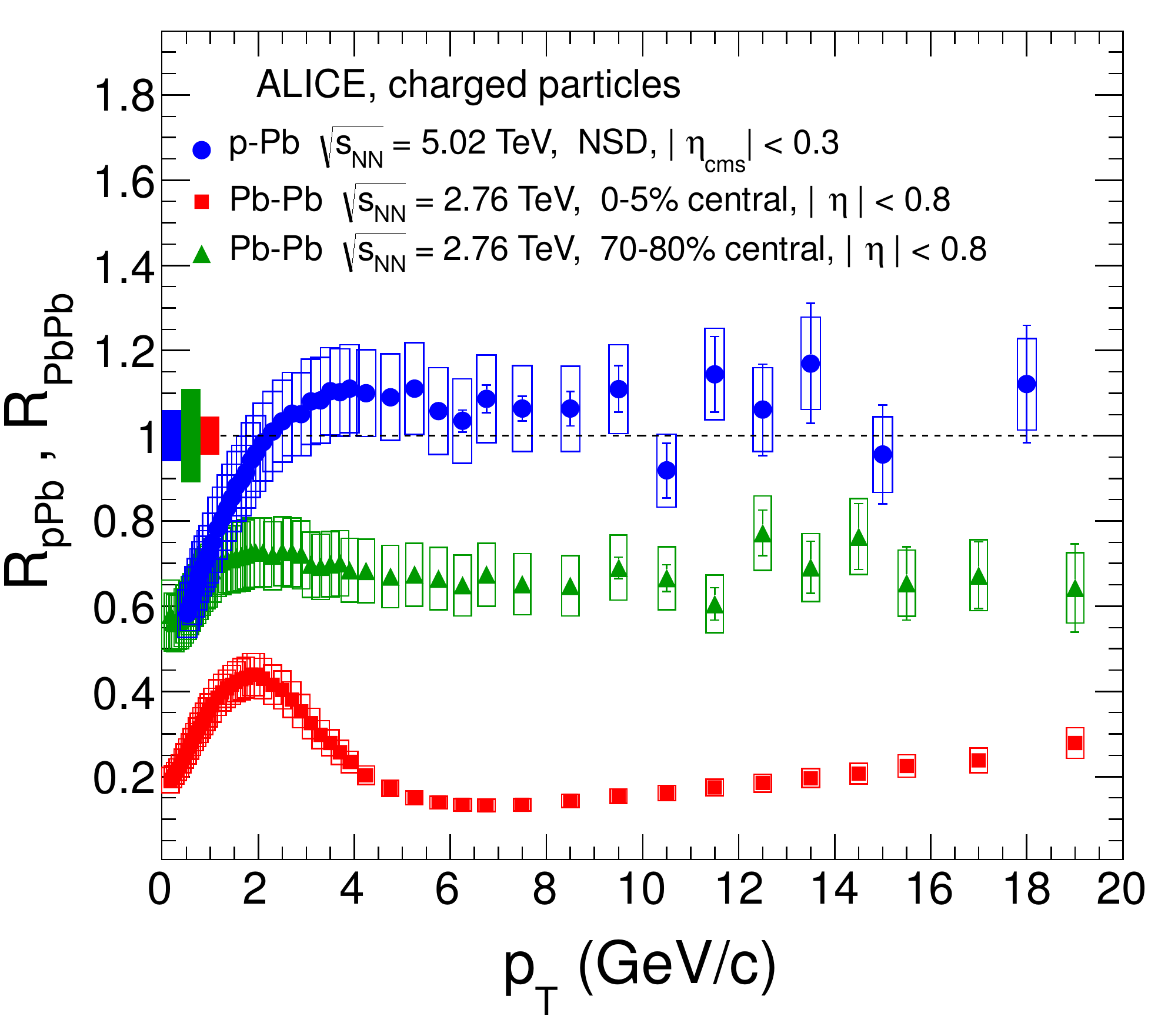}
\includegraphics[width=0.32\textwidth]{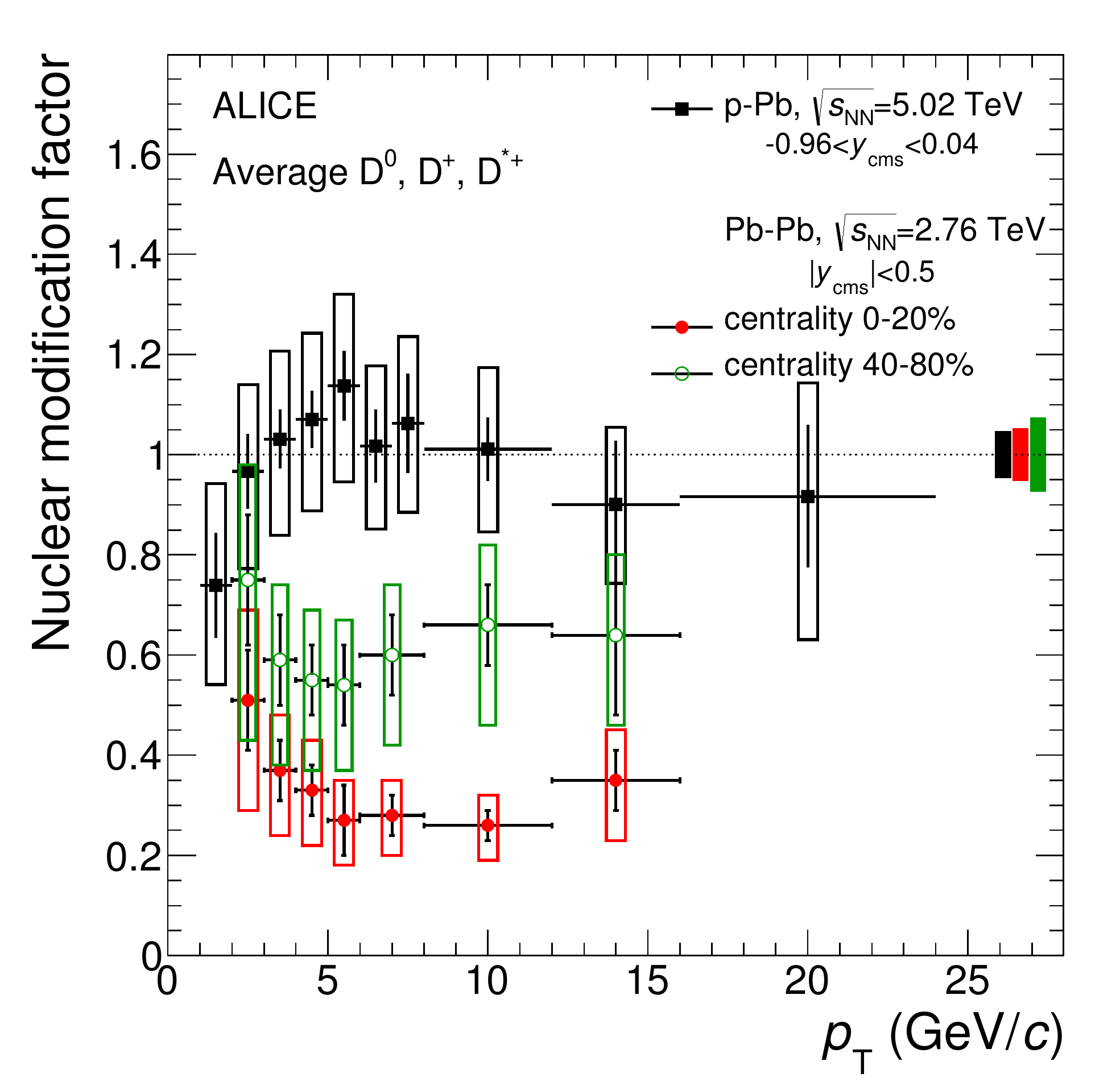}
\vspace{-0.2cm}
\caption[]{\footnotesize {Left: $R_{pPb}$ of charged particles at $\sqrt{s_{NN}}=5.02$ TeV compared to $R_{AA}$ in central (0--5\% centrality) and peripheral (70--80\%) $Pb+Pb$ collisions at $\sqrt{s_{NN}}=2.76$ TeV \cite{RpPb_ALICE_PRL}. Right: Average $R_{pPb}$ of prompt $D^0$, $D^{+}$ and $D^{*+}$ mesons compared to $D$-meson $R_{AA}$ in the 20\% most central and in the 40--80\% $Pb+Pb$ collisions at $\sqrt{s_{NN}}=2.76$ TeV \cite{alice_DpPb} .}} 
\label{RAA_pPb}
\end{figure}

It is now quite well established that at LHC and top RHIC energies QGP is produced in heavy $Pb+Pb$ and $Au+Au$ systems. This is, however, a very important question whether a similar state can be created in collisions of light and intermediate mass systems. Figure~\ref{RAA_pPb} (left) shows the nuclear modification factor for $p+Pb$ collisions at $\sqrt{s_{NN}}=5.02$ TeV compared to $R_{AA}$ in central and peripheral $Pb+Pb$ interactions at $\sqrt{s_{NN}}=2.76$ TeV. One sees that the suppression of high-$p_T$ particles can be seen even in peripheral $Pb+Pb$ collisions, where $\langle N_{coll} \rangle \approx 16$ is only twice higher than $\langle N_{coll} \rangle \approx 7$ in $p+Pb$ interactions (for a comparison in 5\% most central $Pb+Pb$ $\langle N_{coll} \rangle \approx 1700 $). At the top RHIC energy $R_{AA}$ in peripheral $Au+Au$ collisions was close to one. Charged particles in $p+Pb$ collisions at LHC do not show suppression; similar observation was done for $d+Au$ interactions at the top RHIC energy. 
In Fig.~\ref{RAA_pPb} (right) the same conclusions can be drawn from $D$ meson production. Heavy $D$ mesons in $p+Pb$ collisions at LHC also do not show suppression, but they are suppressed both in central and (semi)peripheral $Pb+Pb$ interactions. Figure~\ref{RAA_pPb} confirms that the suppression of high-$p_T$ particles in central, and at LHC also in peripheral $Pb+Pb$ collisions is not due to initial-state effects, but rather due to final state interactions in a hot and dense medium (QGP opaque to energetic partons).

\begin{figure}[h]
\begin{tikzpicture}
    \begin{scope} [xshift=-3.5cm]
    \node {\includegraphics[width=0.38\textwidth]{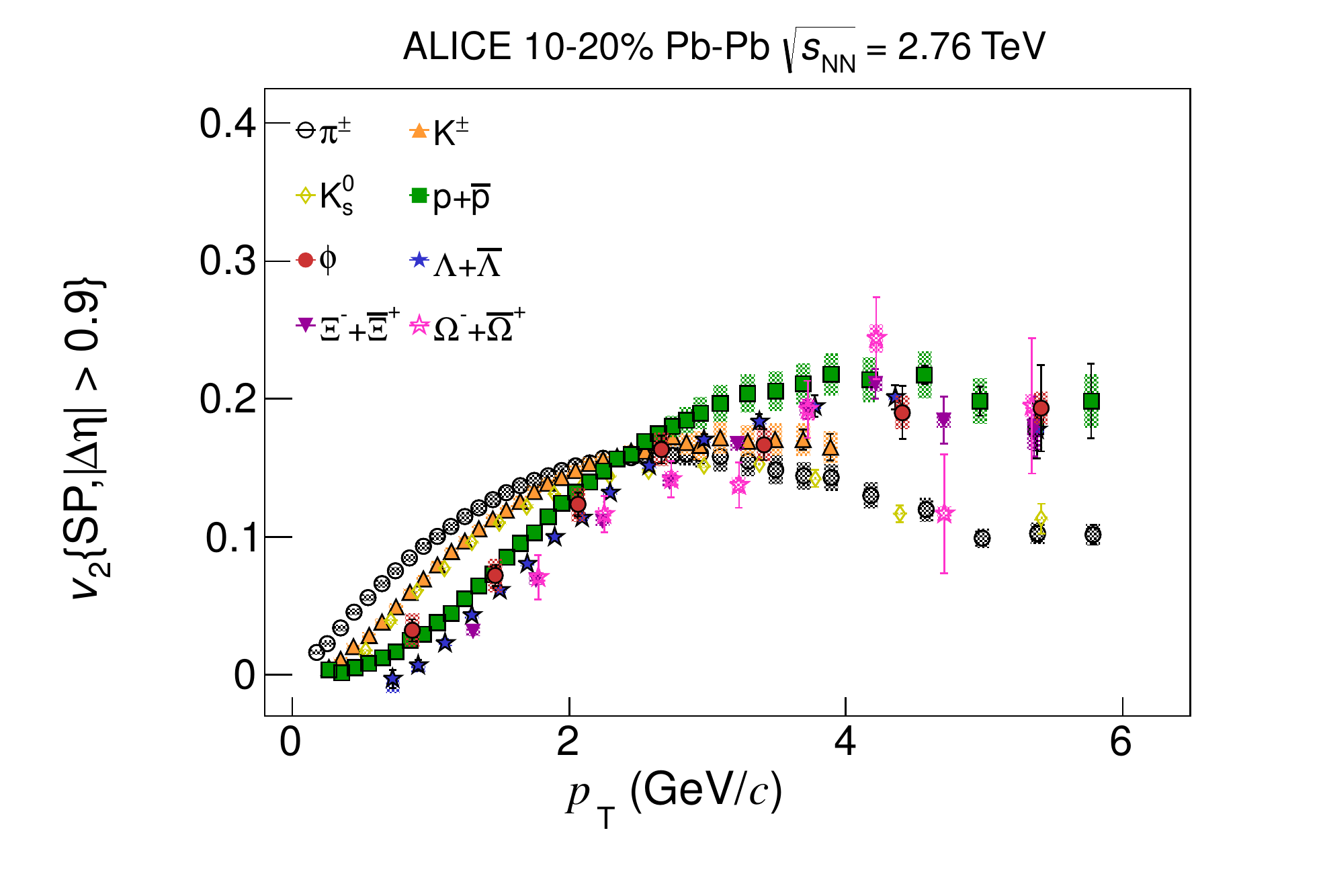}};
    \end{scope}
    \begin{scope} [xshift=1.8cm]
    \node {\includegraphics[width=0.35\textwidth]{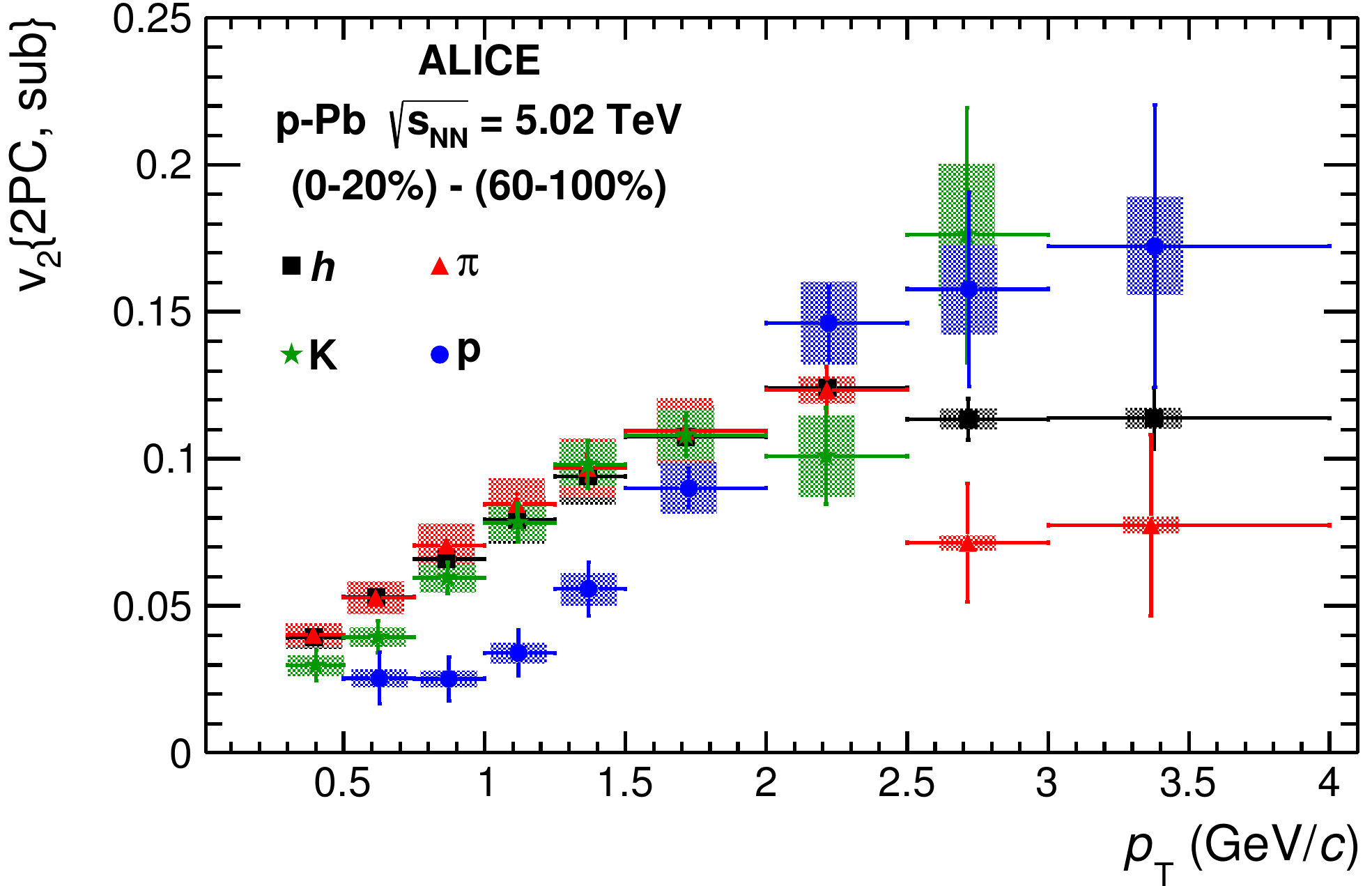}};
    \end{scope}
    \begin{scope} [xshift=6.5cm,  yshift=+0.1cm] 
    \node {\includegraphics[width=0.25\textwidth]{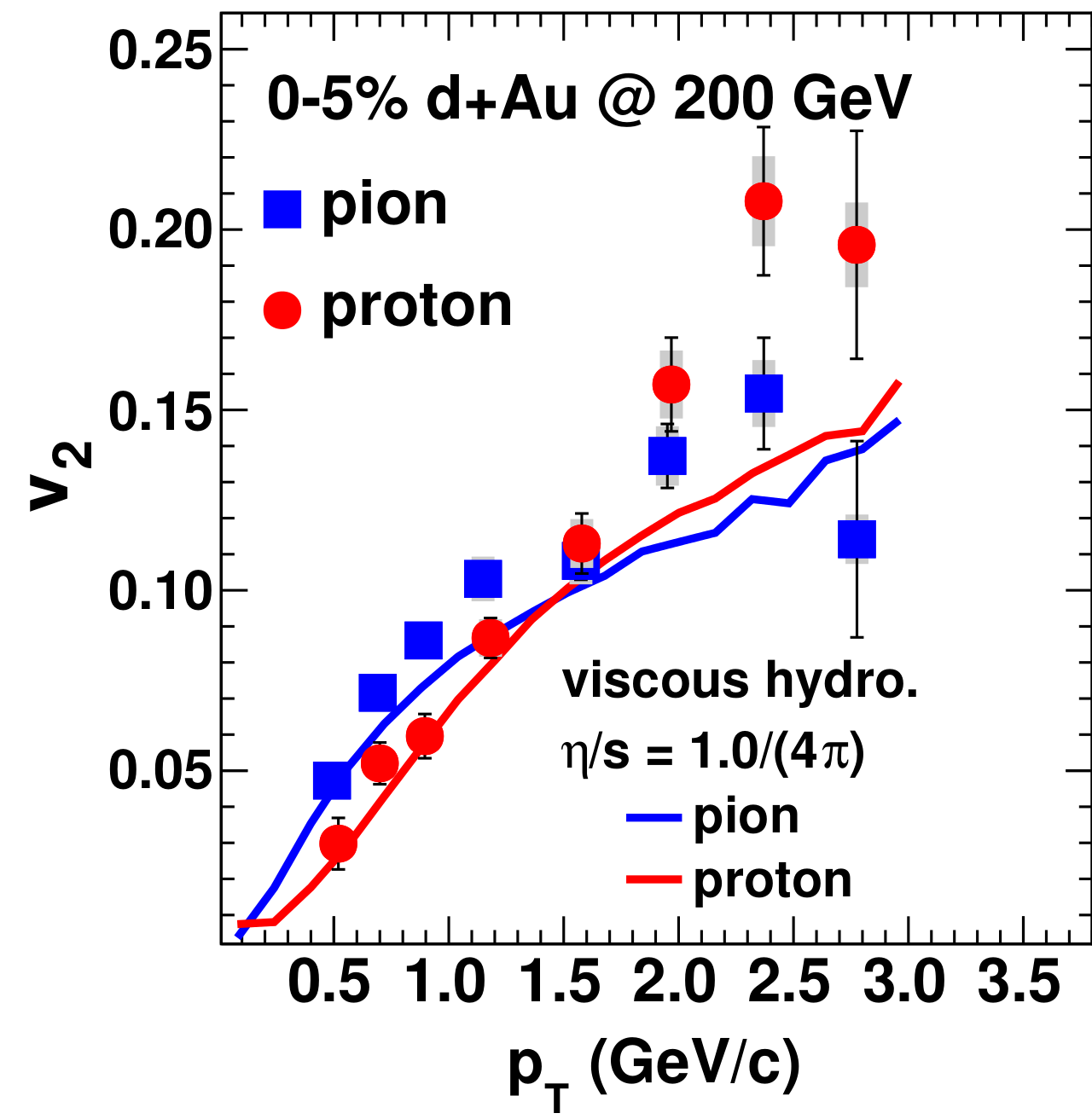}};
    \end{scope} 
\end{tikzpicture}
\vspace{-0.5cm}
\caption[]{\footnotesize {Left: The Fourier coefficients ($v_2$) for different particle species in $Pb+Pb$ collisions at $\sqrt{s_{NN}}=2.76$ TeV \cite{noferini_WPCF2013}. Middle: $v_2$ in $p+Pb$ collisions at $\sqrt{s_{NN}}=5.02$ TeV for hadrons, pions, kaons, and protons from the correlation in the 0--20\% multiplicity class after subtraction of the correlation from the 60--100\% multiplicity class (see \cite{split_v2_pPb} for details). Figure taken from \cite{split_v2_pPb}.  
Right: $v_2(p_T)$ for identified pions and (anti)protons, each charged combined, in 0--5\% central $d+Au$ collisions at $\sqrt{s_{NN}}=200$ GeV \cite{split_v2_dAu_RHIC}. }}
\label{flow_small_sys}
\end{figure}

Minimum bias $p+Pb$ collisions at LHC do not show similarities to heavy $Pb+Pb$ interactions. The situation is, however, different if we look for {\it central} $p+Pb$ interactions. Figure~\ref{flow_small_sys} shows an evidence of the collective flow in small systems. In $Pb+Pb$ (Fig.~\ref{flow_small_sys} (left)) collisions at LHC, and also in $Au+Au$ interactions at the top RHIC energy (not shown), a mass ordering of $v_2$ was observed. Such a hierarchy (for a given $p_T$ $v_2$ is lower for a higher-mass particle) is qualitatively reproduced (at lower $p_T$) by hydrodynamical models (not shown) and understood as due to radial flow. In hydro models the elliptic flow follows: $v_2 \sim (p_T - \langle v_T \rangle m_T)/T$, where $v_T$ is the transverse expansion velocity (radial flow), $m_T$ is a particle transverse mass, and $T$ temperature. Qualitatively a similar mass ordering of $v_2$ has been observed in high multiplicity $p+Pb$ collisions at LHC (Fig.~\ref{flow_small_sys} (middle)), and recently also in central $d+Au$ collisions at the top RHIC energy (Fig.~\ref{flow_small_sys} (right)). Does $p+Pb$ ($d+Au$) indeed flow? The another evidence of the radial flow in $p+Pb$ collisions is an increase of $\langle p_T \rangle$ with increasing particle mass (see \cite{meanpt_vs_mass_pPb}). The same behaviour is well known from $Pb+Pb$ data \cite{alice_BWfits} and can be reproduced by hydrodynamical models, where for radial flow $\langle p_T \rangle _{(m)} \sim m\, v_T$. 
Surprisingly, similar effects have been also observed by the CMS experiment in $p+p$(!) interactions at $\sqrt{s}=$0.9, 2.76, and 7 TeV \cite{CMS_Tslopes_pp} (so far $p+p$ collisions have been treated as elementary interactions, where the formation of any medium was not expected). The flow-like behaviour in LHC high multiplicity $p+Pb$ events (both $v_2$ mass splitting and the increase of $\langle p_T \rangle$ with increasing particle mass) can be reproduced by recent hydrodynamical calculations (see for example \cite{broniowski}). The interaction region is small but dense and perhaps we can apply hydrodynamics to high-multiplicity $p+p$ and $p+A$ collisions.

\begin{figure}[h]
\centering
\includegraphics[width=0.32\textwidth]{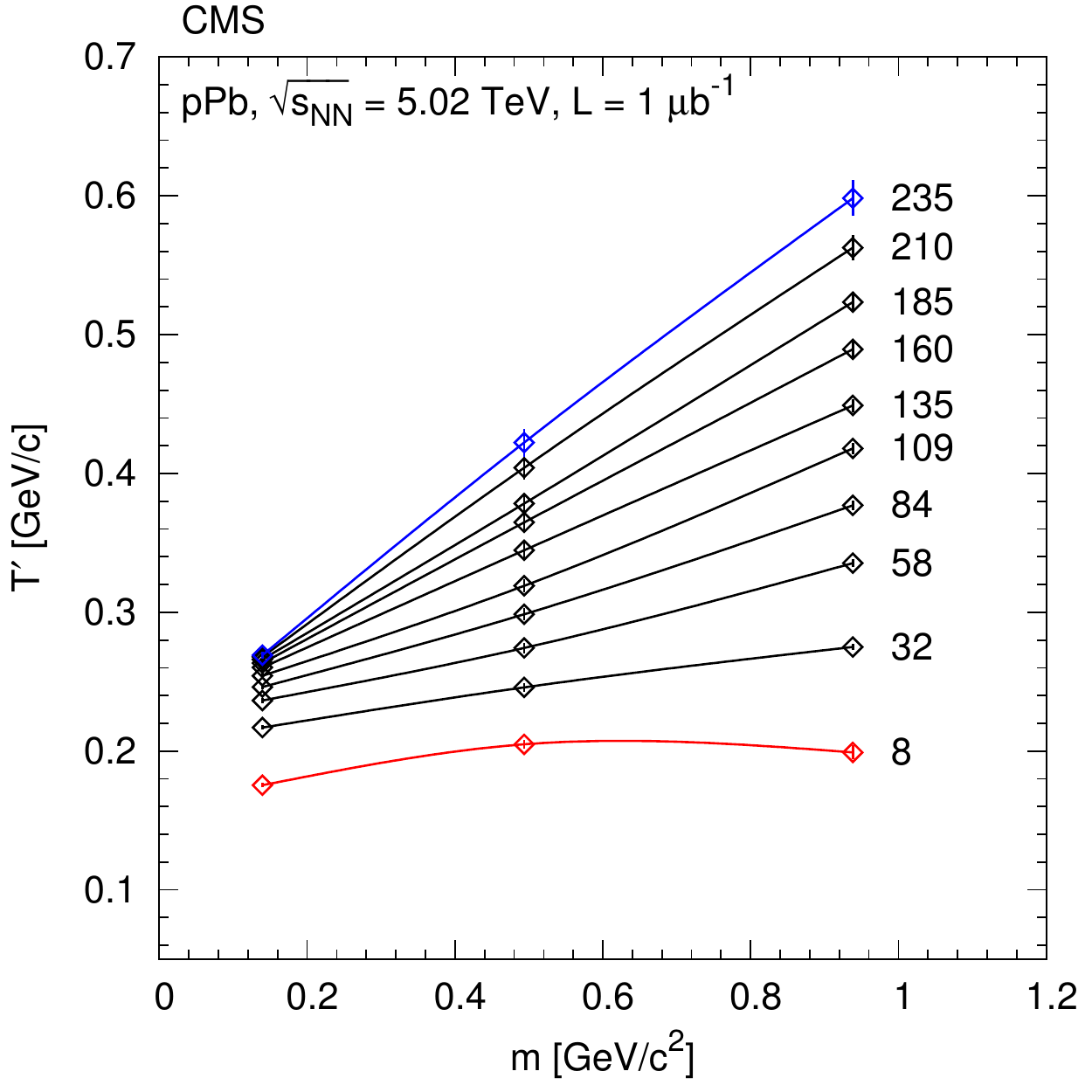}
\includegraphics[width=0.45\textwidth]{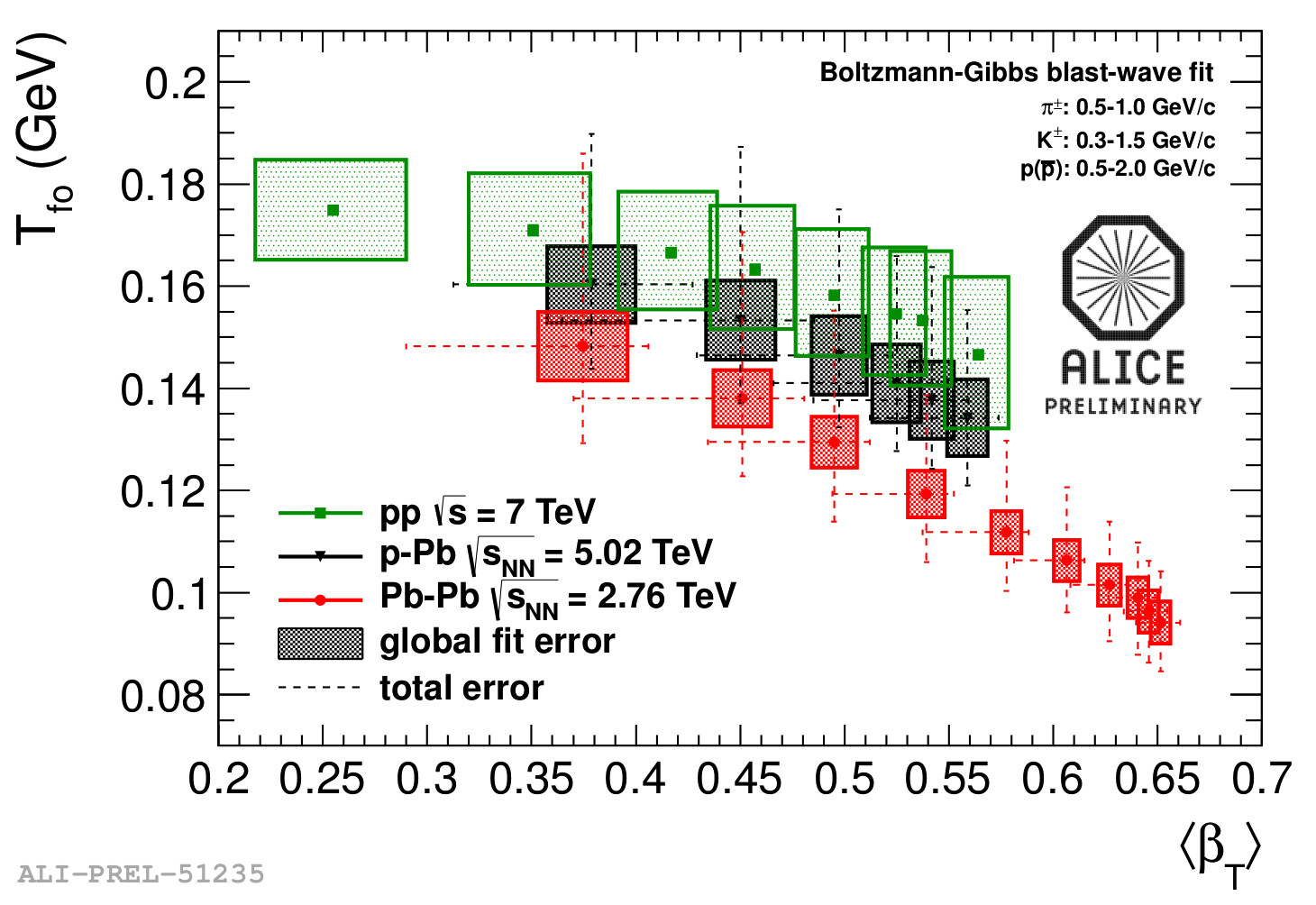}
\vspace{-0.2cm}
\caption[]{\footnotesize {Left: Inverse slope parameters of exponential $p_T$ spectra for different multiplicity bins ($\langle N_{tracks} \rangle$ given as the numbers next to lines) of $p+Pb$ collisions at $\sqrt{s_{NN}}=5.02$ TeV \cite{CMS_Tslopes}. Right: Thermal freeze-out parameters ($T_{fo}$, $\langle \beta_{T} \rangle $) fitted within the Blast-Wave Model \cite{BW_th} to $p+p$, $p+Pb$, and $Pb+Pb$ spectra (different centralities are shown) at LHC energies. Figure taken from \cite{timmins_wwnd14}; see ALICE $Pb+Pb$ results \cite{alice_BWfits} for details. }}      
\label{T_CMS_ALICE_BW}
\end{figure}

The more direct evidence of the radial flow in small systems can be seen in Fig.~\ref{T_CMS_ALICE_BW}. The left panel shows inverse slope parameters ($T$) of exponential $p_T$ spectra (Boltzmann-type distribution: $dN/dp_T \propto p_T \exp (-m_T/T)$) for different multiplicity bins ($\langle N_{tracks} \rangle$ given as the numbers next to the lines) of $p+Pb$ collisions at $\sqrt{s_{NN}}=5.02$ TeV. In the case of a static source the inverse slope parameter ($T$) is equal to the thermal (kinetic) freeze-out temperature ($T_{fo}$). In the case of an expanding source $T$ is equal to the freeze-out temperature ($T_{fo}$) plus the effect of the radial flow. The example for a non-relativistic case ($p_{T, i}  \ll m_i$) follows: $T \approx T_{fo} + \frac{1}{2} m_i \langle v_T \rangle ^2$. Figure~\ref{T_CMS_ALICE_BW} (left) shows that in LHC $p+Pb$ collisions the flow-like behaviour (increase of $T$ with a particle mass) becomes much stronger for the highest multiplicity events.

The thermal freeze-out parameters ($T_{fo}$ and the mean transverse flow velocity $\langle \beta_T \rangle$) can be more precisely obtained via fits within the Blast-Wave Model \cite{BW_th}, where $p_T$ spectra follow: 
\begin{equation}
\frac {1}{p_T} \frac{dN}{dp_T} \propto \int _0 ^R r\, dr\, m_T I_0 \left( \frac{p_T \sinh \rho(r)}{T_{fo}}\right)  K_1 \left( \frac{m_T \cosh \rho(r)}{T_{fo}} \right).
\end{equation}  
$I_0$ and $K_1$ are modified Bessel functions, $\rho(r) = \tanh ^{-1} \beta_T(r)$, and the transverse velocity profile $\beta_T(r) \equiv \beta_{T(surface)} (r/R)^n$, where $R$ is the radius of the fireball\footnote {$\langle \beta_T \rangle = 2/(2+n) \beta_{T(surface)}$, and thus $\langle \beta_T \rangle$ is smaller than $\beta_{T(surface)}$. }. 
Figure~\ref{T_CMS_ALICE_BW} (right) shows the results of the Blast-Wave Model fits for different centralities (or multiplicities) of $Pb+Pb$, $p+Pb$, and $p+p$ collisions at LHC energies. For more central events higher $\langle \beta_T \rangle$ values were obtained. Moreover, a similar evolution of $T_{fo}$ versus $\langle \beta_T \rangle$ is seen in all systems. In the most central $Pb+Pb$ collisions $\langle \beta_T \rangle$=0.65c (10\% higher than in $Au+Au$ at the top RHIC). In central $p+Pb$ interactions $\langle \beta_T \rangle \sim 0.5$c, but similar values are also seen in $p+p$ collisions! Thus, Fig.~\ref{T_CMS_ALICE_BW} (right) should be considered as a possible sign of collectivity in $p+Pb$ and $p+p$ interactions\footnote{Other explanations are also available on the market - see LHC $p+p$ results \cite{CR_ALICE} in the PYTHIA model with Color Reconnection \cite{CR_original} (CR) mechanism, which acts on a microscopic level, and therefore does not require formation of thermalized medium in a small system. Generally, there is no flow in PYTHIA, but CR (color string formation between final partons from independent hard scatterings) can mimic flow-like trends seen in $p+p$ data. Note, that Color Reconnections are coherent effects between strings and therefore they might be treated as a some form of collectivity!}.


\section{SPS and lower RHIC energies: ''boiling water''}

If we want to study the phase transition region and search for the critical point of strongly interacting matter we should focus on the SPS and lower RHIC energies. There are dedicated energy scan programs both at SPS (NA49 and NA61/SHINE experiments) and at RHIC (Beam Energy Scan (BES) program with STAR and PHENIX experiments).

\vspace{0.5cm}
\subsection{Looking for the onset of deconfinement energy}

\begin{figure}[h]
\centering
\vspace{0.3cm}
\includegraphics[width=0.4\textwidth]{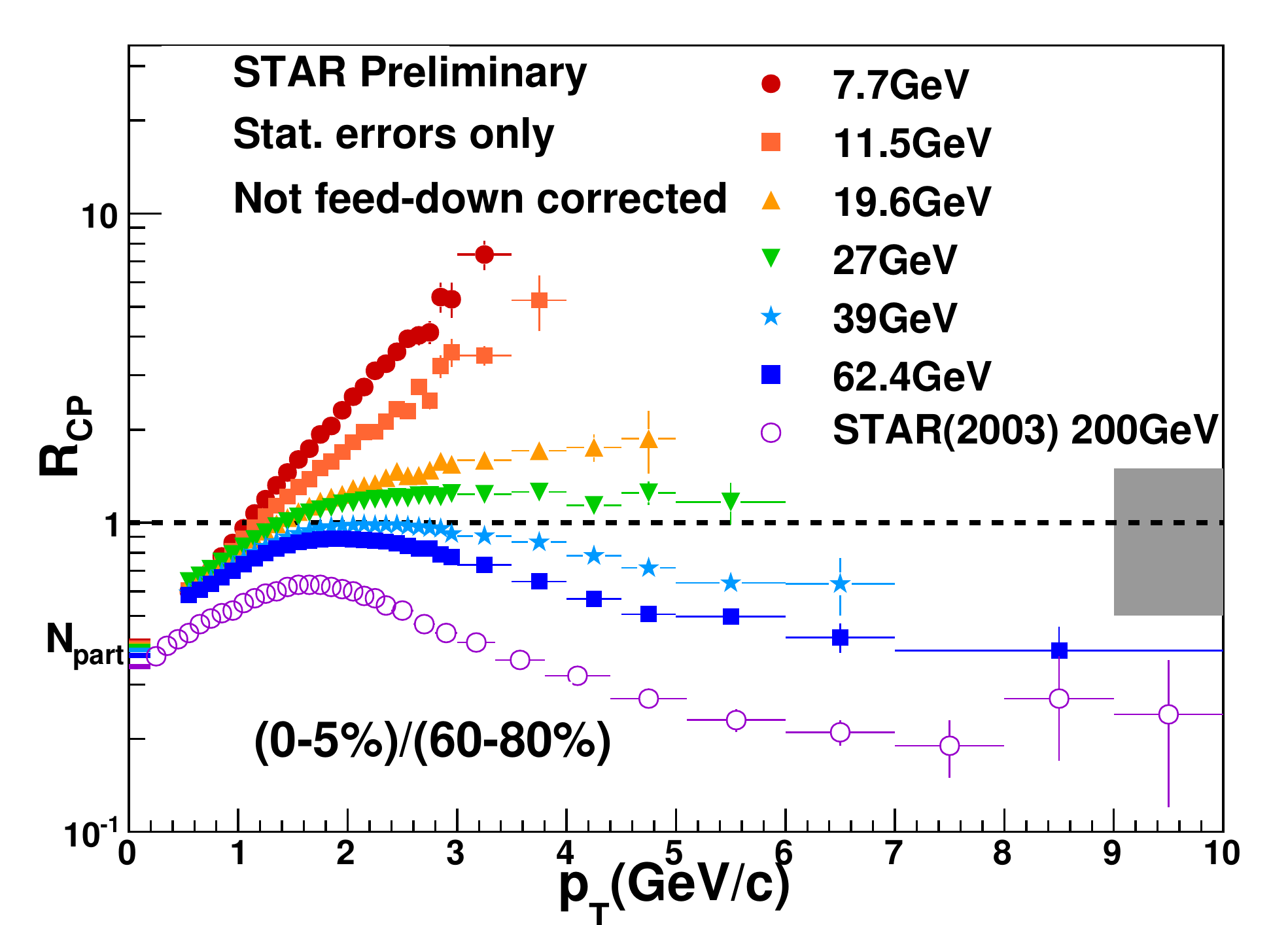}
\includegraphics[width=0.52\textwidth]{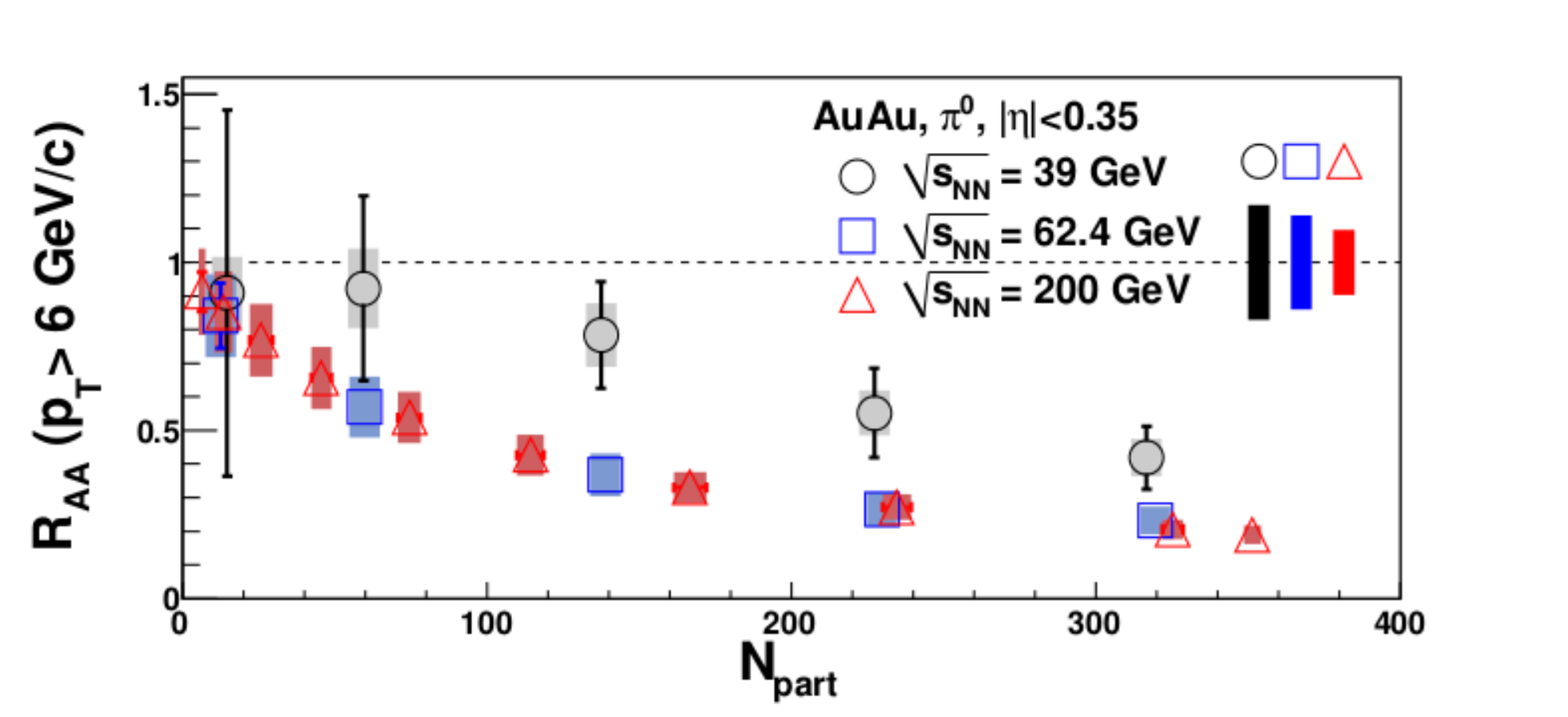}
\vspace{-0.2cm}
\caption[]{\footnotesize {Left: STAR preliminary results on the energy dependence of charged particles $R_{CP}$ \cite{RCP_STAR_BES}. 
Right: $R_{AA}$ for neutral pions with $p_T >6$ GeV/c for $\sqrt{s_{NN}}=$ 200, 62.4, and 39 GeV $Au+Au$ collisions as a function of the number of participants (measure of centrality; high $N_{part}$ represents most central collisions) \cite{Mitchel_RAA_BES}. }}
\label{raa_bes_star_phenix}
\end{figure}

Figure~\ref{raa_bes_star_phenix} (left) shows the energy dependence of charged particles $R_{CP}$ measured in $Au+Au$ collisions within STAR BES program. As seen, jet quenching disappear at lower energies (absence of a dense medium), and a figure of this type is quite often referred to as a {\it ''turn-off'' of QGP signature}. At lower energies partonic effects become less important and cold nuclear matter effects (Cronin) start to dominate (see also HIJING results with jet quenching off \cite{RCP_BES_HIJING}). Figure~\ref{raa_bes_star_phenix} (right) presents $R_{AA}$ measured by PHENIX for neutral pions with $p_T >6$ GeV/c at three BES energies. The suppression of high-$p_T$ particles is similar at $\sqrt{s_{NN}}=200$ GeV and $\sqrt{s_{NN}}=62.4$ GeV, but significantly smaller at $\sqrt{s_{NN}}=39$ GeV.

\begin{wrapfigure}{r}{8.cm}
\vspace{-0.7cm}
\centering
\includegraphics[width=0.42\textwidth]{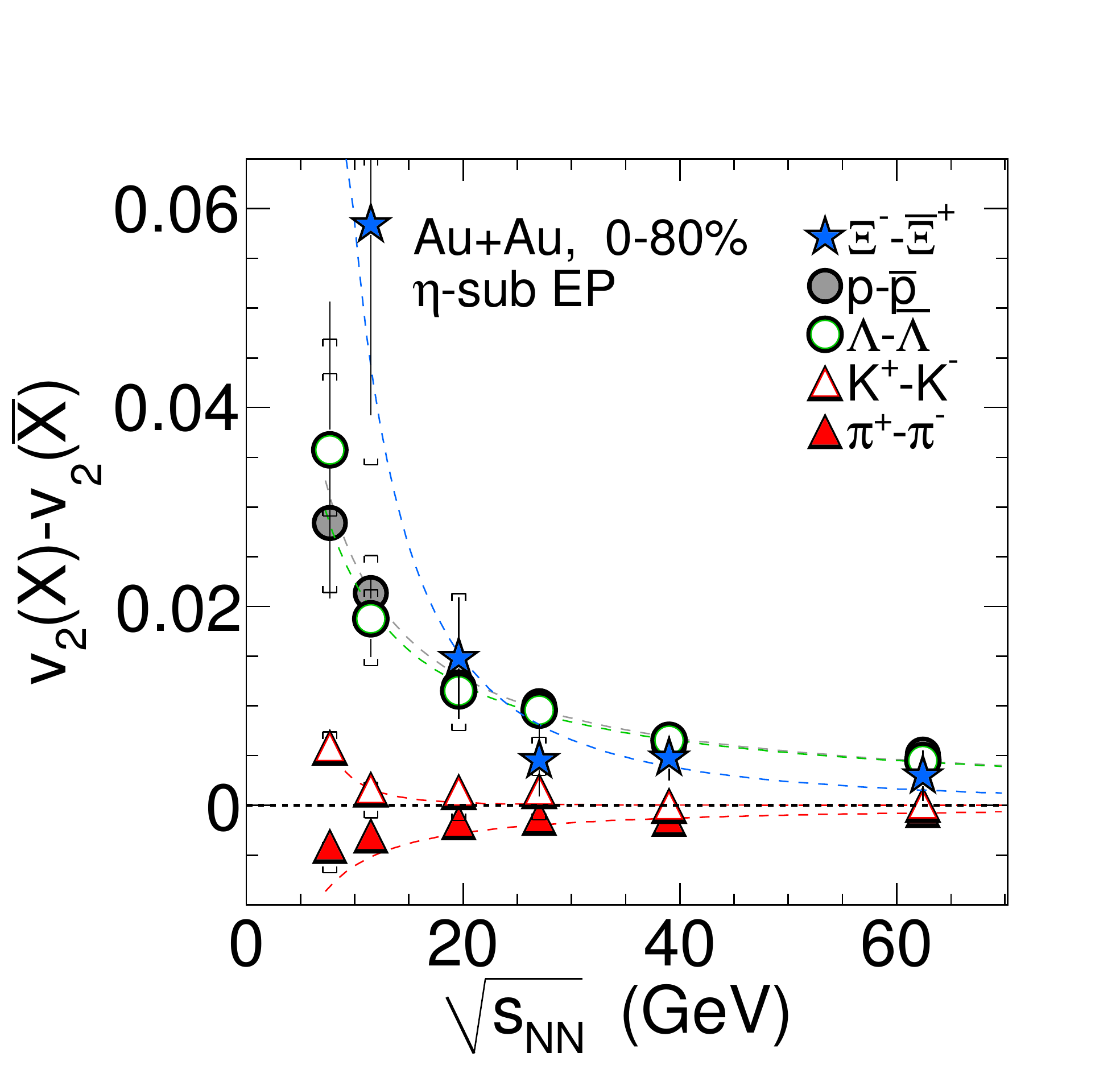}
\vspace{-0.2cm}
\caption[]{\footnotesize {The difference in $v_2$ between particles ($X$) and their corresponding anti-particles ($\bar{X}$) as a function of energy for 0--80\% central $Au+Au$ collisions. 
Figure taken from \cite{v2scal_break_BES}.}}
\label{v2part_antipart_STAR}
\end{wrapfigure}

The observation of breaking of the NCQ scaling at lower energies can be also treated as a new method of estimation of the onset of deconfinement energy. This type of {\it ''turn-off'' of QGP signature} was also observed by the STAR experiment within BES program. The scaling of the elliptic flow with $n_q$ favors partonic degrees of freedom, whereas breaking of $v_2$ scaling with $n_q$ favors hadronic degrees of freedom. Figure~\ref{v2part_antipart_STAR} shows the difference between $v_2$ of the particle and its antiparticle at lower energies. Breaking of the partonic collectivity at lower energies may be interpreted as a change of degrees of freedom in the system (departing from QGP region).  

\clearpage

\begin{figure}[h]
\centering
\includegraphics[width=0.35\textwidth]{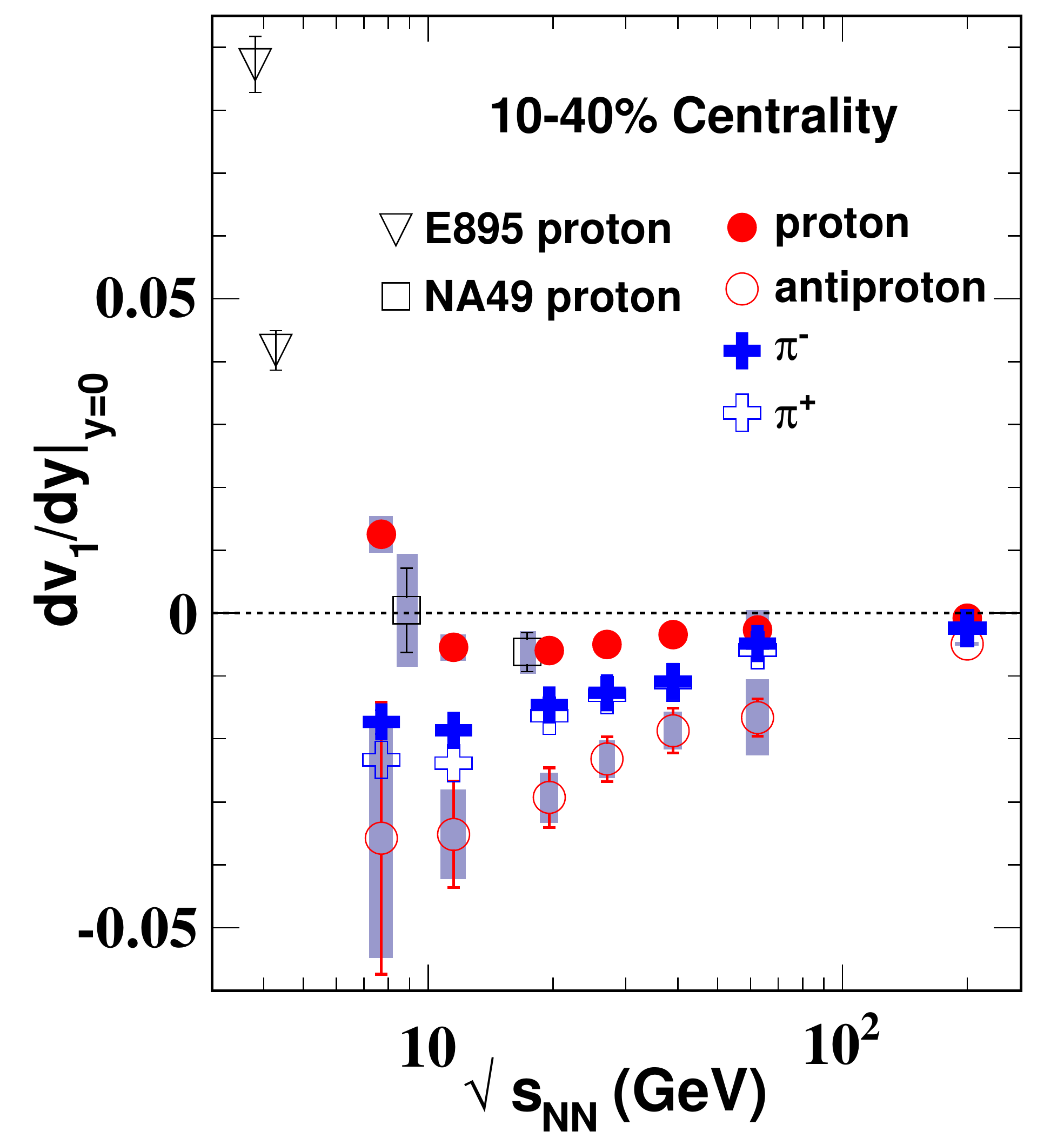}
\includegraphics[width=0.33\textwidth]{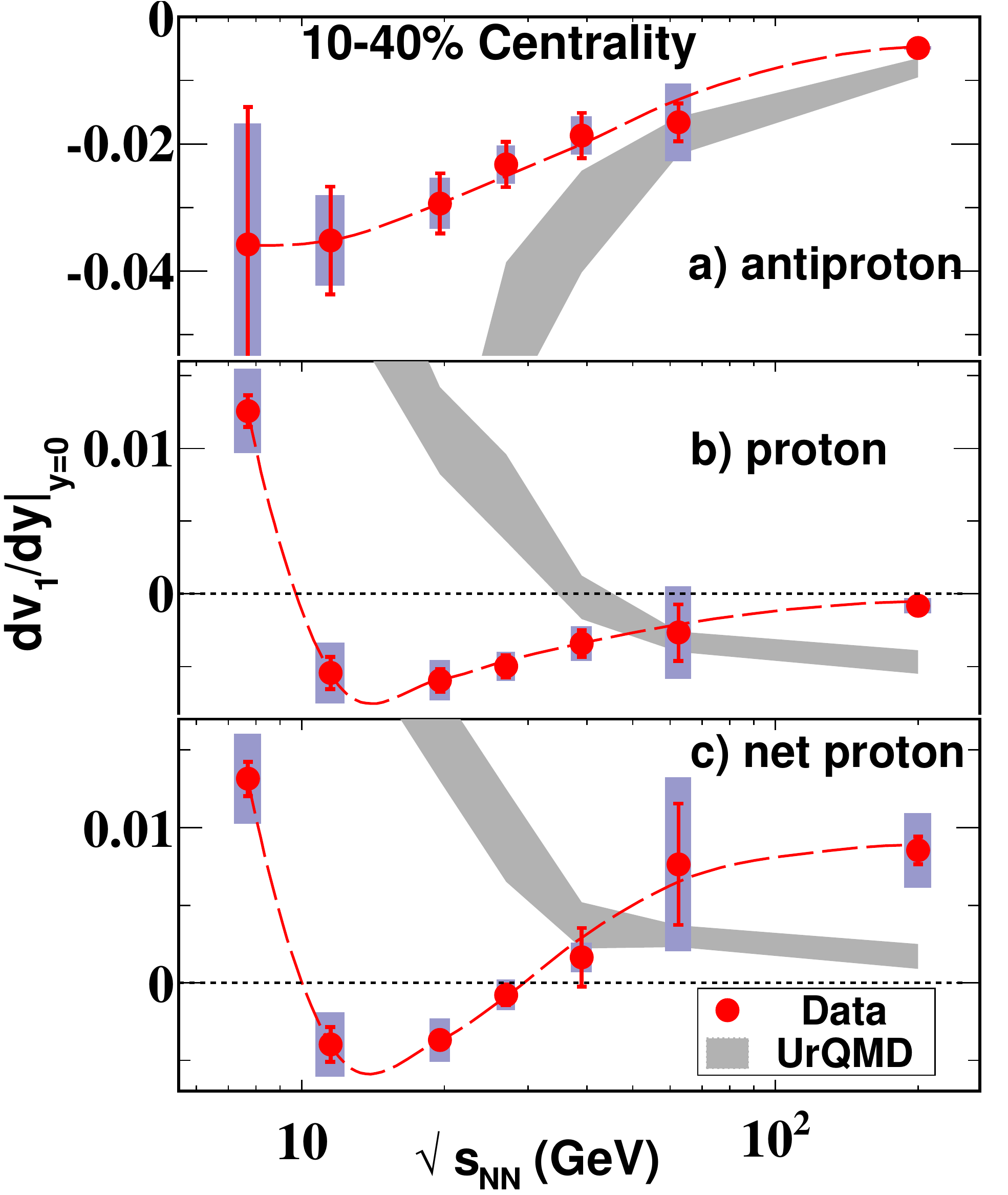}
\vspace{-0.2cm}
\caption[]{\footnotesize {Directed flow slopes ($dv_1/dy$) near mid-rapidity versus energy for intermediate-centrality $Au+Au$ collisions. In the right panel results are compared to prediction of the UrQMD model. 
Figure taken from \cite{v1slope_STARBES}. }}
\label{v1_slopes}
\end{figure}

The next signature connected with the energy threshold for deconfinement is based on a measurement of the directed flow as a function of rapidity\footnote {Rapidity $y=\frac{1}{2} \ln \frac{E+p_L}{E-p_L}$, where $p_L$ is a particle longitudinal momentum (beam direction) and $E$ its total energy.}. Directed flow ($v_1$) was considered to be sensitive to the 1st order phase transition (strong softening of the Equation of State) \cite{v1_softening_EOS}. In principle, the ''collapse of proton flow'', namely, the non-monotonic behaviour (positive $\rightarrow$ negative $\rightarrow$ positive) of proton $dv_1/dy$ as a function of beam energy was expected. The recent STAR results for $Au+Au$ collisions are shown in Fig.~\ref{v1_slopes}. The $v_1$ slopes are always negative for pions and anti-protons, whereas $v_1$ slopes for protons and net-protons change signs at lower energies and show minimum at 10--20 GeV. The additional data sets at $\sqrt{s_{NN}}$=15 GeV are expected to be recorded by STAR in the next phase of BES program (BES-II). This may allow to determine the position of the minimum more precisely. The results shown in Fig.~\ref{v1_slopes} are consistent with hydrodynamical models assuming 1st order phase transition from hadron gas to QGP \cite{v1_softening_EOS}, however, some calculations suggest that the scenario with a cross-over type transition may lead to similar structures \cite{v1_softening_EOS_crossover}.

The Beam Energy Scan program at RHIC collected data which allowed to roughly estimate the energy threshold for deconfinement. Two of the three signatures shown above are based on the observation of the disappearance of a given QGP signal seen previously at higher energies ({\it ''turn-off'' of QGP signature}). But we can estimate the energy threshold for deconfinement even more precisely. This was done several years ago by the NA49 experiment at CERN SPS, where ''sharp onset'' of deconfinement was observed. 
In 2002 the NA49 experiment completed the SPS energy scan of central $Pb+Pb$ collisions. This program was originally motivated by predictions of the Statistical Model of the Early Stage (SMES) \cite{mg_model} assuming that the energy threshold for deconfinement (the lowest energy sufficient to create a partonic system) is located at low SPS energies. Several structures were expected within SMES: the {\it kink} in pion production (due to increased entropy production), the {\it horn} in the strangeness to entropy ratio, and the {\it step} in the inverse slope parameter of transverse mass spectra (constant temperature and pressure in a mixed phase). Such signatures were indeed observed in $A+A$ collisions by the NA49 experiment \cite{na49pikp}, thus locating the onset of deconfinement (OD) energy around 30$A$ GeV beam energy ($\sqrt{s_{NN}} \approx 7.6$~GeV). Until recently the evidence of OD was based on the results of a single experiment. Last years the results on central $Pb+Pb$ collisions at the LHC \cite{lhc} and data on central $Au+Au$ collisions from the RHIC BES program \cite{bes} were released. They allowed to verify the NA49 results and their interpretation by STAR and ALICE.

\begin{figure}
\centering
\includegraphics[width=0.32\textwidth]{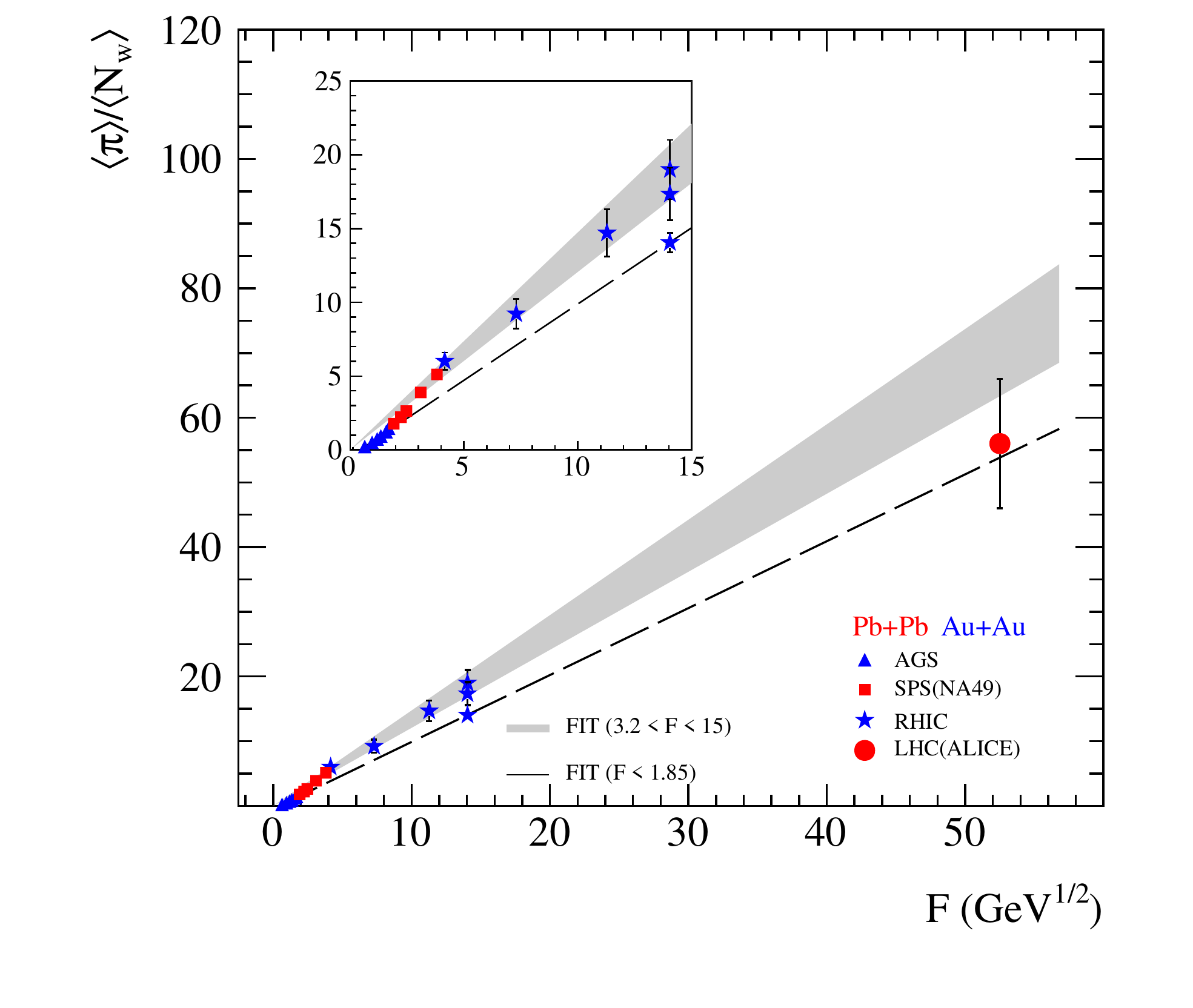}
\includegraphics[width=0.32\textwidth]{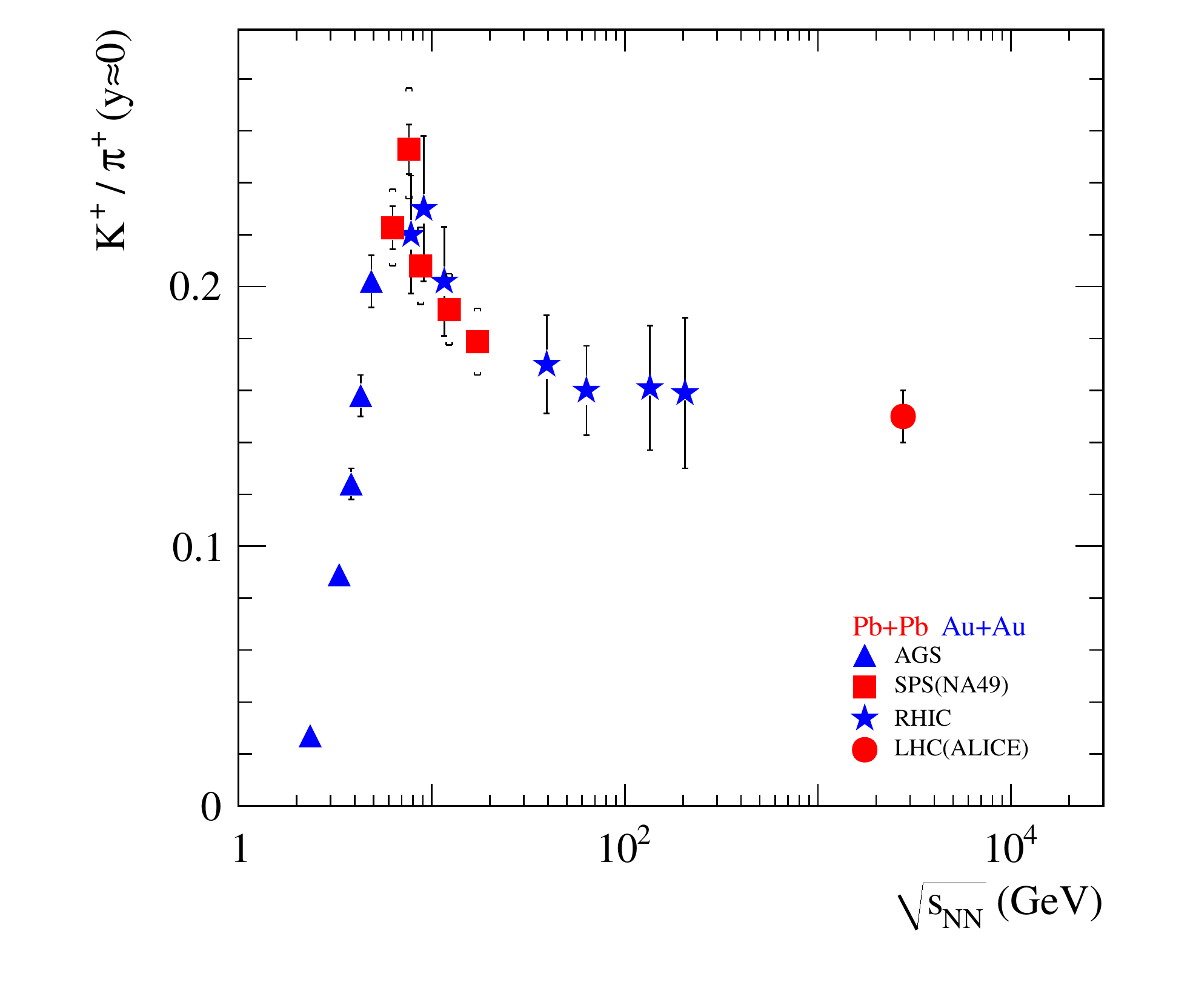}
\includegraphics[width=0.32\textwidth]{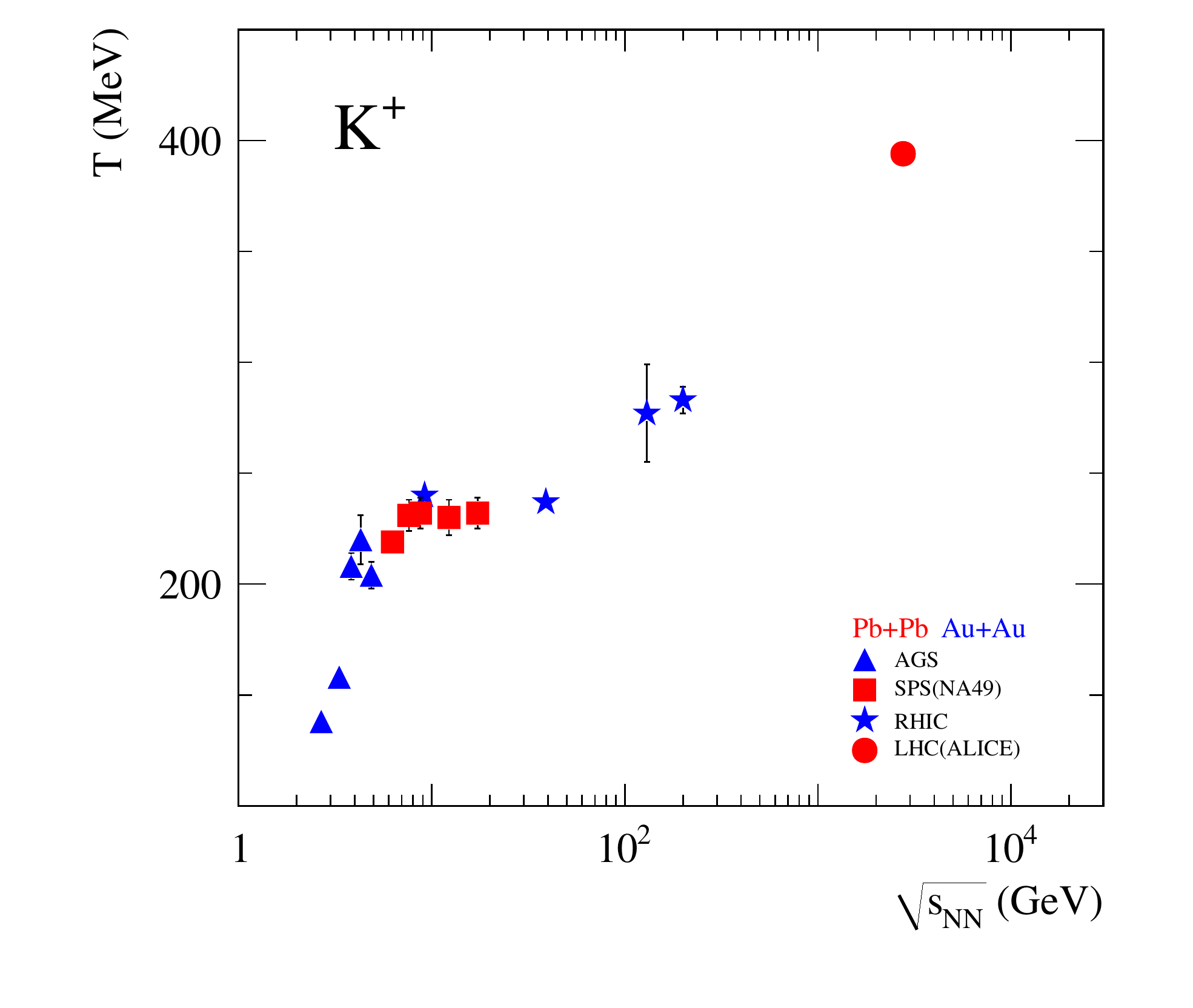}
\vspace{-0.5cm}
\caption[]{\footnotesize {Left: Energy ($F$) dependence of the mean pion multiplicity per 
wounded nucleon in full phase space ($4\pi$). Middle: Energy dependence of the $\langle K^{+} \rangle / \langle \pi^{+} \rangle $ ratio near mid-rapidity. Right: Energy dependence of the inverse slope parameter ($T$) of the transverse mass spectra of $K^{+}$ mesons. Figures taken from \cite{anar}.}}
\label{kink_horn_step}
\end{figure}

Figure~\ref{kink_horn_step} (left) shows production of charged pions (the total entropy is 
carried mainly by pions\footnote{In SMES the total entropy and the total 
strangeness are the same before and after hadronization (the entropy cannot 
decrease during the transition from QGP to hadron gas), therefore pions 
measure the early stage entropy.}) $\langle \pi \rangle = 1.5 (\langle \pi^{+} 
\rangle + \langle \pi^{-} \rangle) $ normalized to the number of 
wounded nucleons versus Fermi variable $F$ ($F \approx 
(s_{NN})^{1/4}$). In SMES, this ratio is proportional to the 
effective number of degrees of freedom ($NDF$) to the power of 1/4.
For central $A+A$ collisions ($Pb+Pb$ for SPS and LHC or $Au+Au$ for AGS and RHIC) a 
change of slope around 30$A$~GeV ($\sqrt{s_{NN}}=7.6$ GeV) is visible (slope in $A+A$ increases from 
$\approx$ 1 (AGS) to $\approx$ 1.3 (top SPS+RHIC) - consistent with 
increase by a factor of 3 in $NDF$). Such an increase is not observed for
$p + p (\bar{p})$ reactions (not shown). The increase in $NDF$, when going from hadron 
gas to QGP, may be interpreted as a consequence of the activation of partonic 
degrees of freedom. The RHIC BES points follow the line for $A+A$ 
collisions and the LHC point\footnote{The mean pion multiplicity at LHC 
was estimated based on the ALICE measurement of charged particle 
multiplicity, see \cite{anar} for details.}, within a large error, does 
not contradict extrapolations from high SPS and RHIC energies.

Figure~\ref{kink_horn_step} (middle) presents the $\langle K^{+} \rangle / \langle \pi^{+} 
\rangle $ ratio near mid-rapidity versus energy. In SMES, the ratio is proportional to strangeness/entropy densities. Results for $A+A$ are very different from the results for $p+p$ ($p+p$ are not shown here; see \cite{na49pikp}) and show a sharp peak ({\it horn}) in $\langle K^{+} \rangle / \langle \pi^{+} \rangle $ at $\sqrt{s_{NN}} \approx 7.6$~GeV. As seen, RHIC results confirm NA49 measurements at the onset of deconfinement. Moreover, LHC (ALICE) data demonstrate that the energy dependence of hadron production properties shows rapid changes only at low SPS energies, and a smooth evolution is observed between the top SPS ($\sqrt{s_{NN}}$=17.3 GeV) and the current LHC ($\sqrt{s_{NN}}$=2.76 TeV) energies. 
The measure which much better reflects the total strangeness to entropy 
ratio in the SPS energy range is $E_s = (\langle K \rangle + \langle \Lambda \rangle)/ \langle 
\pi \rangle$, proposed in \cite{mg_model}, and calculated from $\pi$, 
$K$, and $\Lambda$ yields in $4\pi$ acceptance. The $E_s$ ratio can be directly and quantitatively compared to SMES predictions. The $E_s$ ratio shows a distinct peak in $E_s$ at 30$A$ GeV (not shown; see figure in \cite{na49pikp}). This behavior is described (predicted) only by the model assuming a phase transition (i.e. SMES), where the maximum ({\it horn}), is the result of the decrease of strangeness carrier masses in the QGP ($m_s < m_{\Lambda, K, ...}$) and the change in the number of degrees of freedom when reaching the deconfined state.

Figure~\ref{kink_horn_step} (right) presents inverse slope parameters ($T$) of transverse mass spectra\footnote{$m_T$ spectra were parametrized by $dn/(m_T dm_T) = C \cdot exp (-m_T / T)$ and fits were done close to mid-rapidity.} of positively charged kaons. For $A+A$ data one can see a 
strong rise at AGS, plateau at SPS, and rise towards RHIC and LHC energies. Such structure is not observed for $p+p$ collisions (not shown). The plateau is consistent with constant temperature and pressure in the mixed phase (latent heat) \cite{smes_temp}. In fact, this structure ({\it step}) strongly resembles the behavior of water, where a plateau can be observed in the temperature when heat is added. Models (not shown) without phase transition do not reproduce the $A+A$ data, but a hydrodynamical model incorporating a deconfinement phase transition at SPS energies 
\cite{smes_hydro} describes the results in Fig.~\ref{kink_horn_step} (right) quite well. 

All three structures ({\it kink}, {\it horn}, {\it step}) confirm that results agree with the interpretation of the NA49 structures as due to OD. Above the onset energy only a 
smooth change of QGP properties with increasing energy is expected.

\begin{wrapfigure}{r}{8.cm}
\centering
\vspace{-0.9cm}
\includegraphics[width=0.5\textwidth]{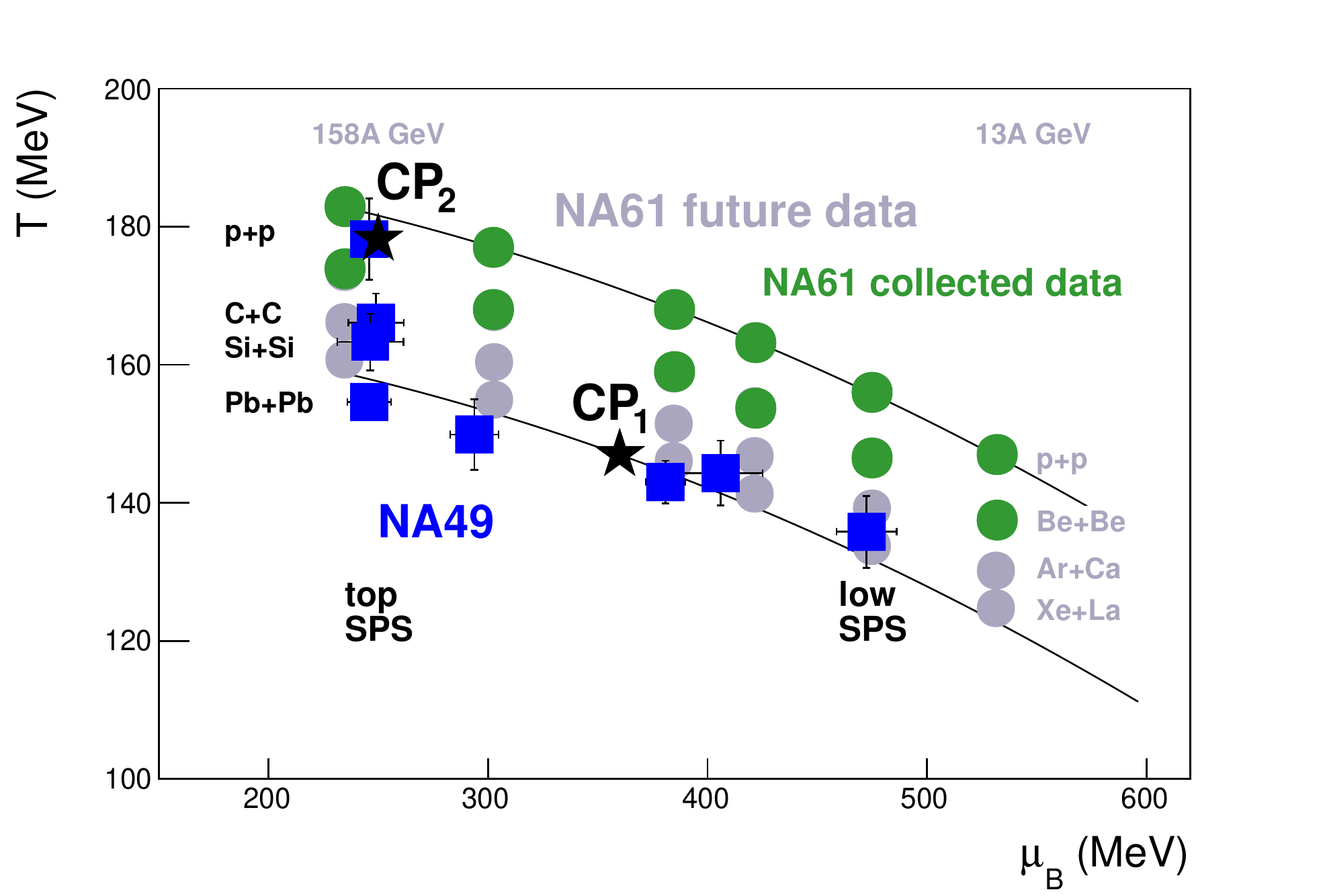}
\vspace{-0.2cm}
\caption[]{\footnotesize {Estimated (NA49) and expected (NA61) chemical 
freeze-out points according to \cite{beccatini} and parameterizations therein.
$CP_{1}$ and $CP_{2}$ (see next section) are the points that were considered in NA49 as possible locations of the critical point: $CP_1$ with $\mu_B$ from lattice QCD calculations \cite{fodor_latt_2004} and $T$ on the empirical freeze-out line; $CP_2$ as the chemical freeze-out point of $p+p$ reactions at 158$A$ GeV (assuming that this freeze-out point may be located on the phase transition line).
}}
\label{phase_diagram}
\end{wrapfigure}

\vspace{0.5cm}
The signatures of the onset of deconfinement energy are seen at middle SPS energies for  
heavy $Pb+Pb$ and $Au+Au$ systems. This is, however, a very important question whether the onset of deconfinement can be also seen in collisions of light and intermediate mass systems. This question can be answer by the NA61/SHINE \footnote{SHINE $-$ SPS Heavy Ion and Neutrino Experiment.} 
experiment which is the successor of NA49 (the main detector components are inherited from NA49 but several important upgrades were done). The main goals of the NA61 {\it ion program} are: search for the critical point, study of the properties of the onset of deconfinement, and study high $p_T$ physics (energy dependence of the nuclear modification factor). These goals will be achieved by performing a 
comprehensive scan in the whole SPS energy range ($p_{beam}= 13A-158A$ GeV; $\sqrt{s_{NN}}=5.1-17.3$ GeV) with light and intermediate mass nuclei ($p$, $Be$, $Ar$, $Xe$). It will allow to cover a broad range of the phase diagram (the expected chemical freeze-out points are shown in Fig.~\ref{phase_diagram}), and to search for the onset of the {\it horn}, {\it kink}, {\it step}, etc. in collisions of light nuclei (the structures observed for $Pb+Pb$/$Au+Au$ should vanish with decreasing system size).

\begin{figure}[h]
\begin{tikzpicture}
    \begin{scope} [xshift=-2cm]
    \node {\includegraphics[width=0.325\textwidth]{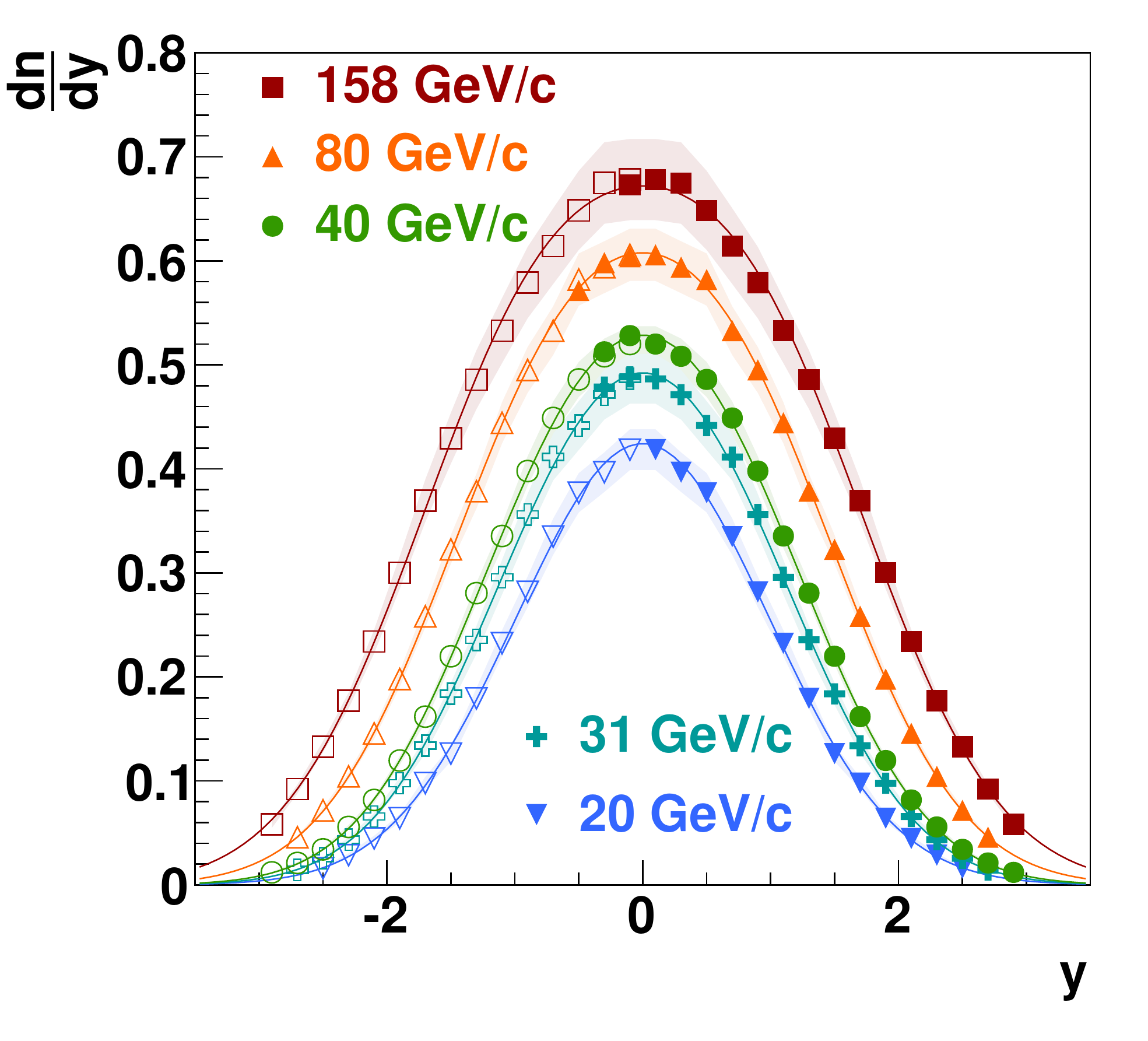}};
    \end{scope}
    \begin{scope} [xshift=3cm]
    \node {\includegraphics[width=0.325\textwidth]{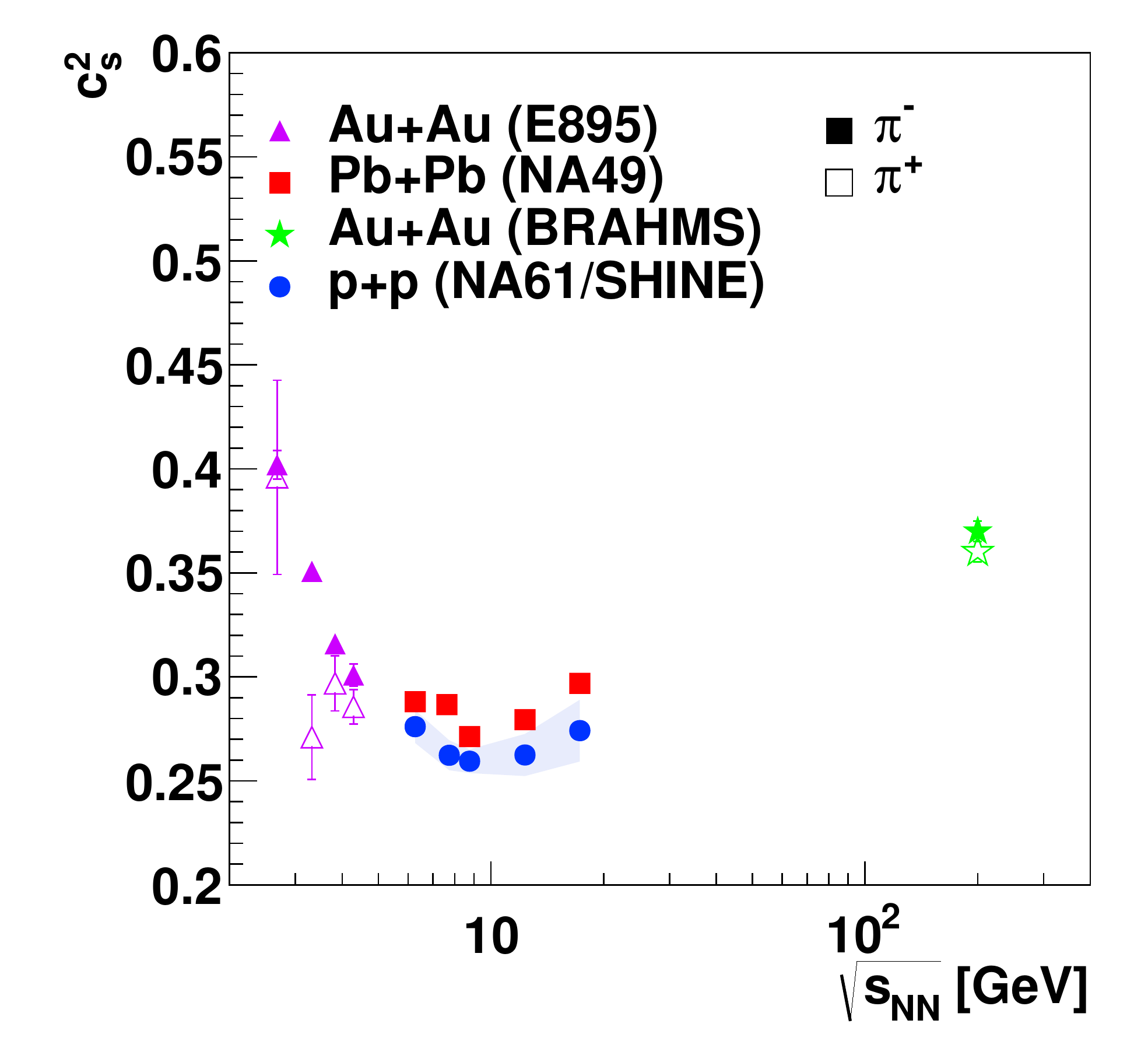}};
    \end{scope}
    \begin{scope} [xshift=8cm, yshift=-1.38cm] 
    \node {\includegraphics[width=0.33\textwidth]{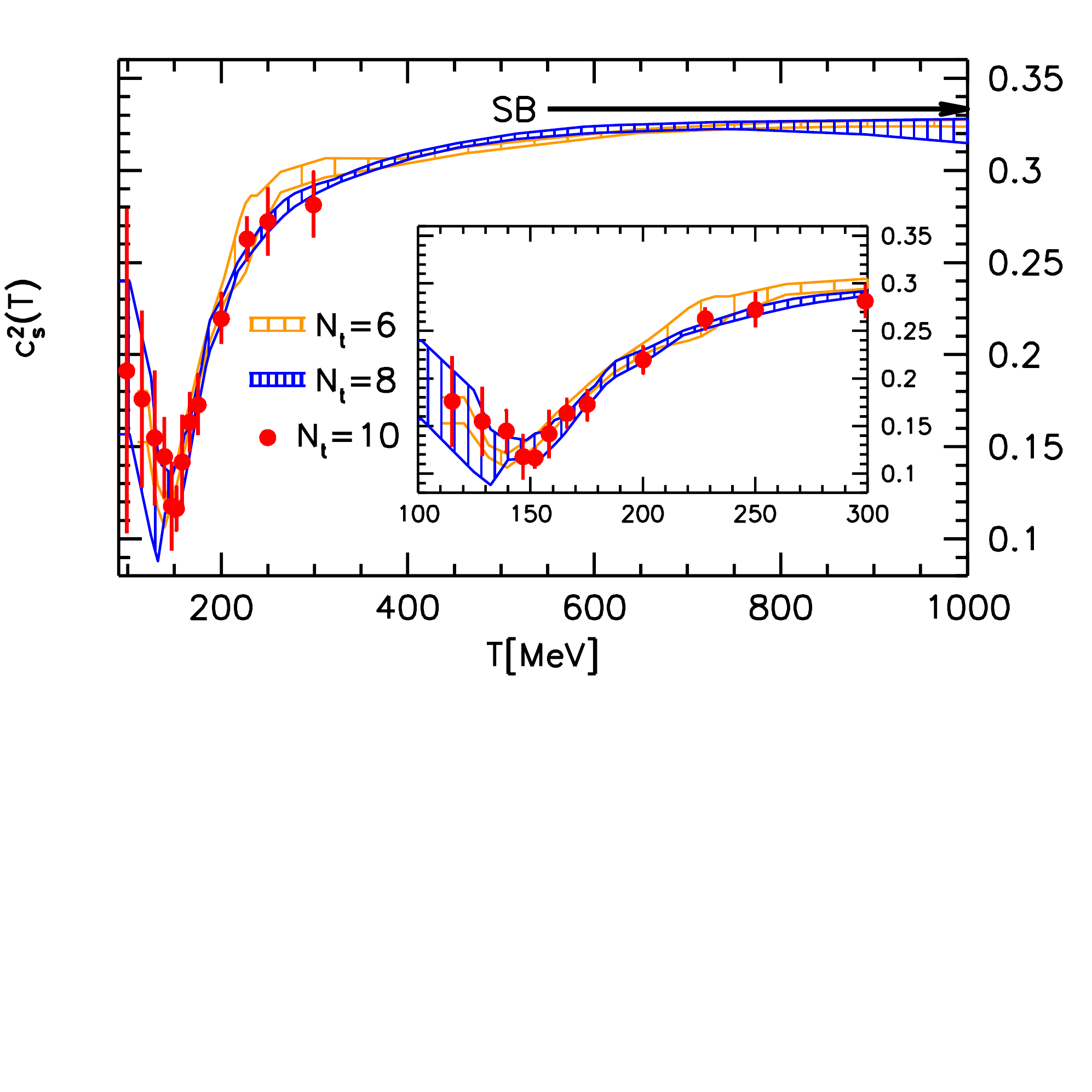}};
    \end{scope} 
\end{tikzpicture}
\vspace{-2.5cm}
\caption[]{\footnotesize {Left: NA61 $p+p$ $m_T$-integrated $\pi^-$ rapidity spectra fitted with a sum of two symmetrically displaced Gaussian functions \cite{antoni_paper}. Middle: Energy dependence of sound velocities obtained from widths of $\pi^{-}$ rapidity spectra (compilation done by A. Aduszkiewicz from NA61). Right: Sound velocity versus temperature obtained from lattice calculations of QCD \cite{latticecs}. }}
\label{dale_fig}
\end{figure}

The first NA61 results on spectra, yields, fluctuations, and correlations in $p+p$ collisions are already available \cite{antoni_paper, cpod_kg, cpod_sp, cpod_mmp} and the {\it kink} and {\it step} plots with NA61 $p+p$ results can be found i.e. in \cite{cpod_kg}. Here we would like to focus on the structure which was not yet discussed above, it is on the {\it dale} \cite{PS_MG_OoD}.   
According to hydrodynamical model \cite{cs_eq} the sound velocity ($c_s$) is related to the width of the rapidity distribution ($\sigma_y$) of pions:
\begin{equation}
\sigma^2_y (\pi^{-}) = \frac {8} {3} \frac {c^2_s} {1-c^4_s} ln (\sqrt{s_{NN}}/2m_p).
\end{equation}
Figure~\ref{dale_fig} (middle) shows the sound velocities obtained from the widths of $\pi^{-}$ rapidity spectra presented in Fig.~\ref{dale_fig} (left). Lattice QCD calculations \cite{latticecs} (see Fig.~\ref{dale_fig} (right)) suggest that the minimum of the sound velocity (softest point of the Equation of State) can be attributed to the phase transition between hadron gas and QGP. Surprisingly, in Fig.~\ref{dale_fig} (middle) such a minimum ({\it dale}) can be seen not only for $Pb+Pb$ collisions but also in $p+p$ data!

\begin{wrapfigure}{r}{8.cm}
\centering
\includegraphics[width=0.35\textwidth]{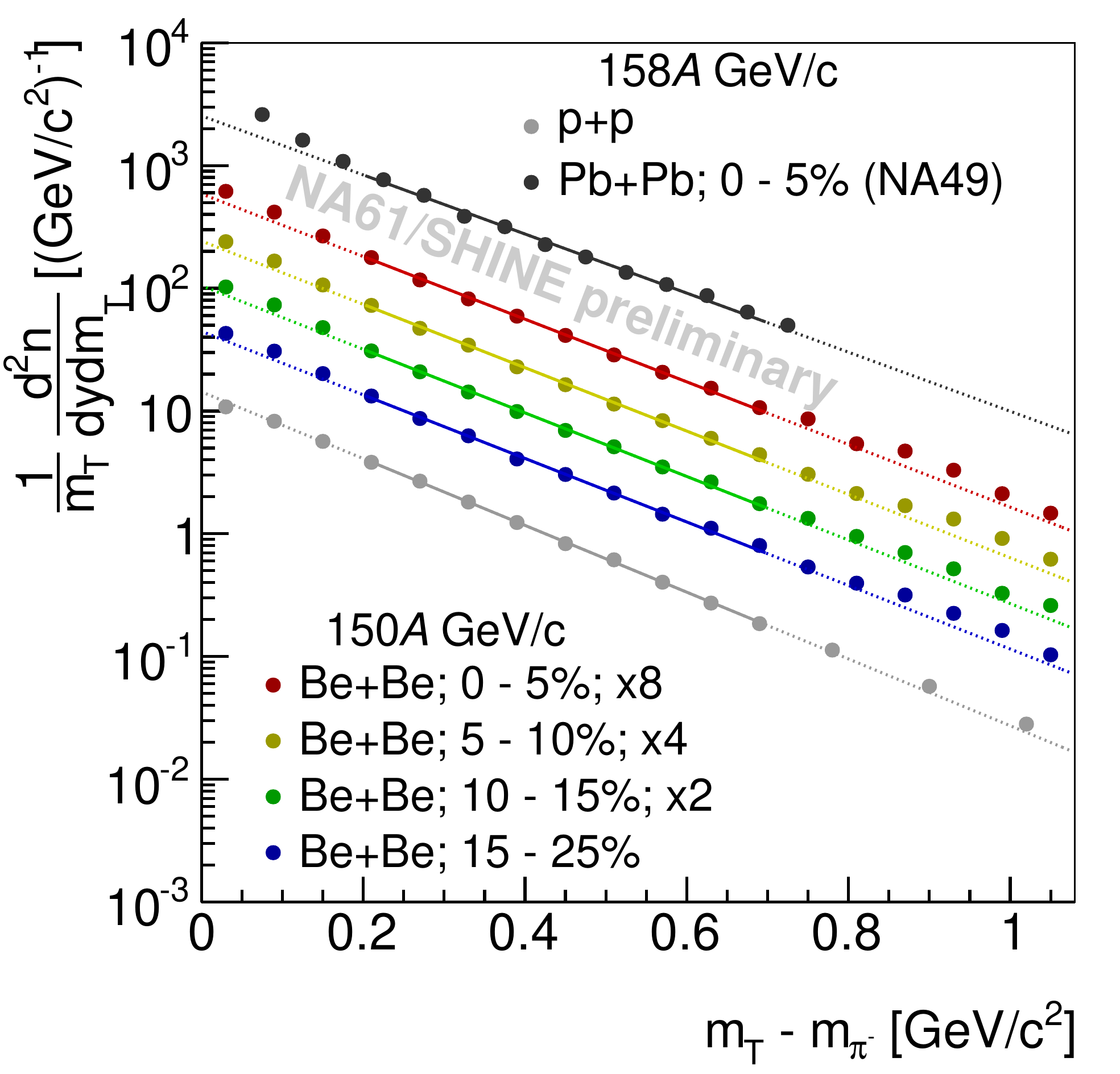}
\vspace{-0.2cm}
\caption[]{\footnotesize {Transverse mass spectra of $\pi^{-}$ in $p+p$ \cite{antoni_paper}, $Be+Be$ \cite{emil_2013} and $Pb+Pb$ \cite{na49pikp} collisions at the top SPS energy. Results are for mid-rapidity $0.0 < y < 0.2$. Figure taken from \cite{emil_2013}.}}
\label{mt_BeBe}
\end{wrapfigure}

\begin{figure}[h]
\centering
\includegraphics[width=0.32\textwidth]{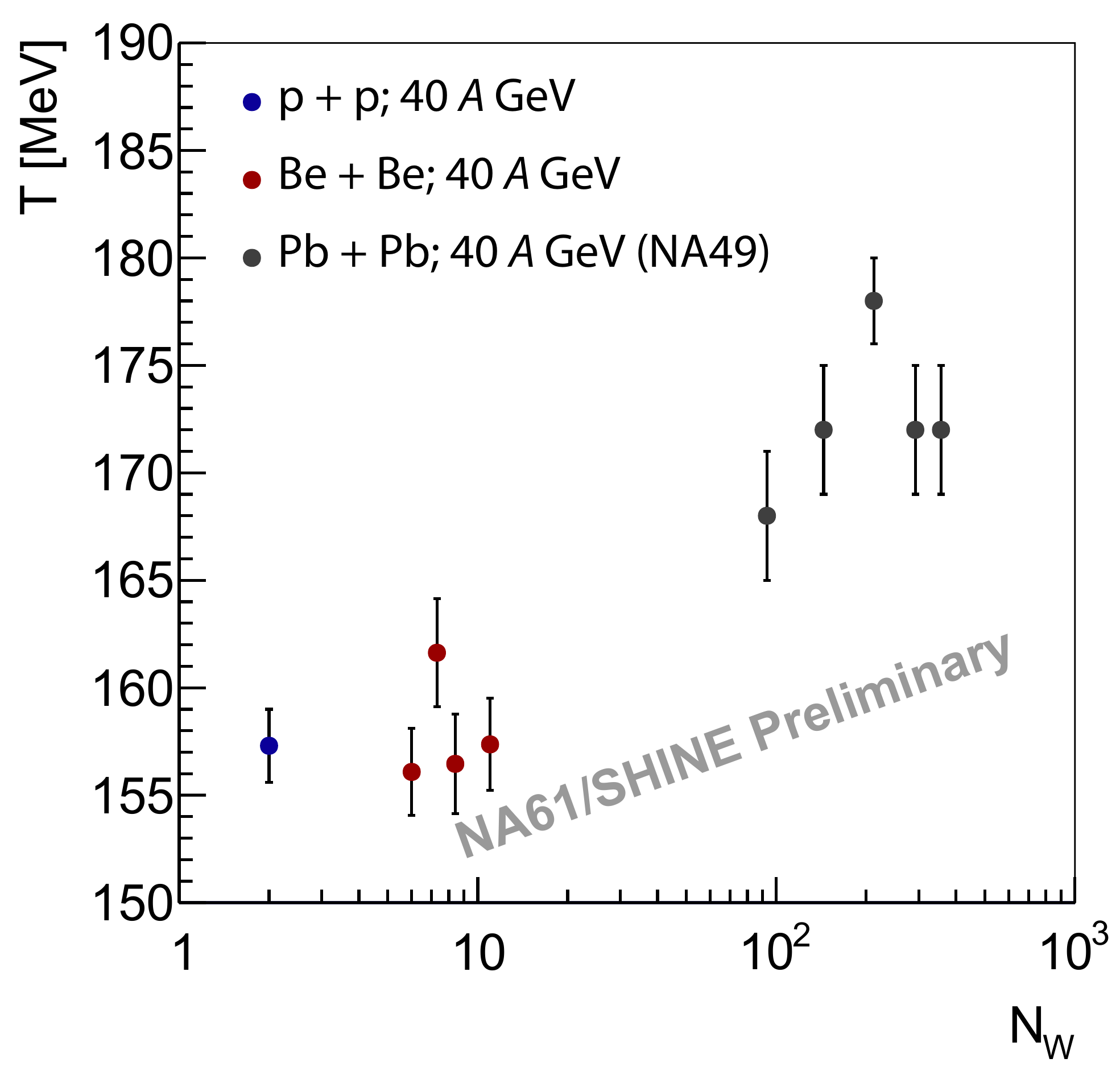}
\includegraphics[width=0.32\textwidth]{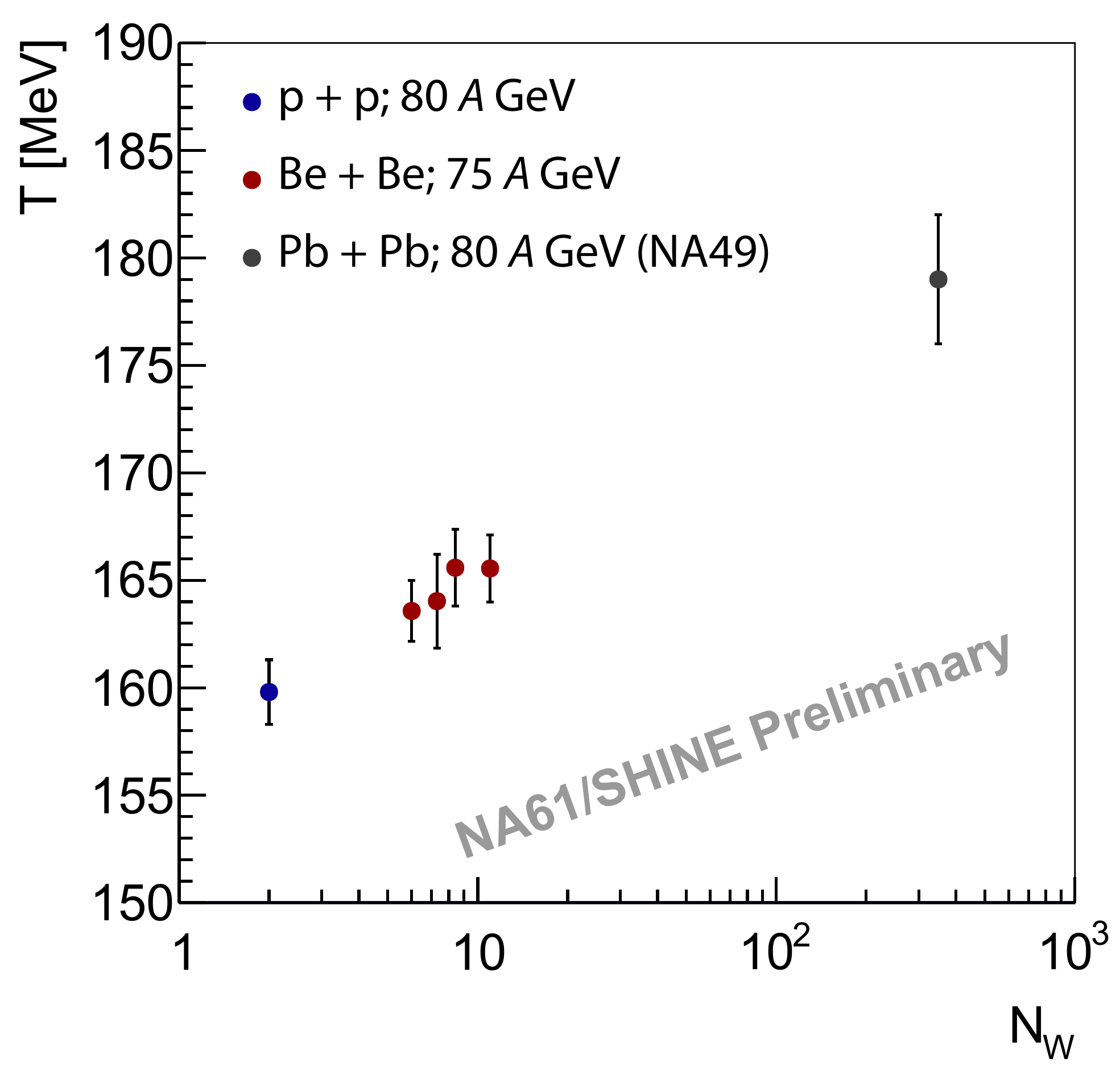}
\includegraphics[width=0.32\textwidth]{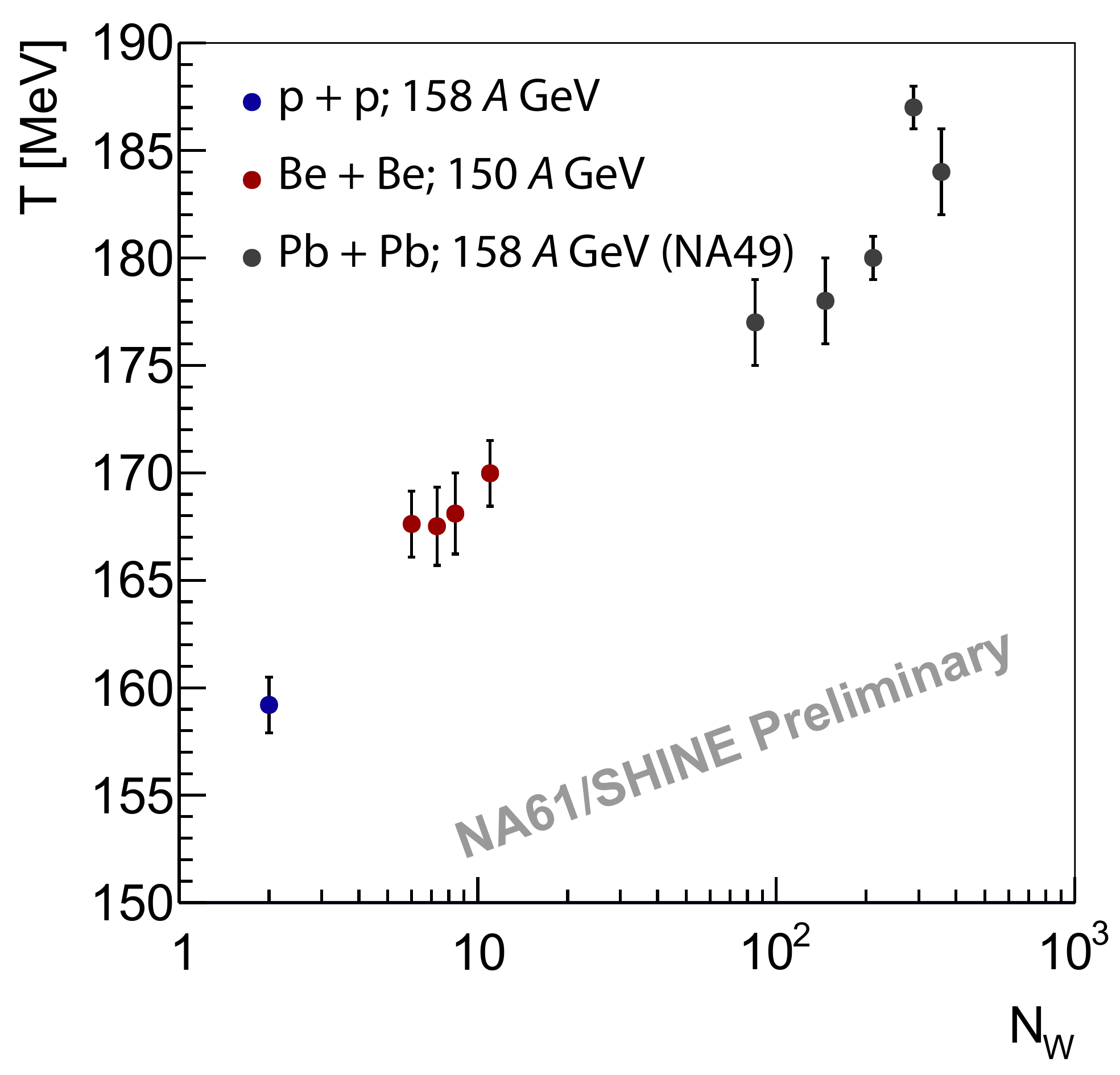}
\vspace{-0.3cm}
\caption[]{\footnotesize {Inverse slope parameters of $m_T$ spectra of $\pi^{-}$ mesons versus number of wounded nucleons \cite{emil_2013}. }}
\label{T_BeBe}
\end{figure}

The first preliminary NA61 results from ion beams ($Be+Be$ collisions) are also available \cite{emil_2013}. When we consider the density of the solid state Beryllium is quite light (1.85 g/cm$^3$) when compared to Lead (11.3 g/cm$^3$). However, at the highest SPS energies the system created after $Be+Be$ collision seems to show the effects which were originally attributed to heavy ion interactions.     
Figure \ref{mt_BeBe} shows mid-rapidity transverse mass spectra of $\pi^{-}$ in $p+p$ \cite{antoni_paper}, $Be+Be$ \cite{emil_2013}, and $Pb+Pb$ \cite{na49pikp} collisions at the top SPS energy. The spectra in $p+p$ collisions are approximately exponential, whereas the convex shape can be seen in $Pb+Pb$ and $Be+Be$ interactions. The spectra were fitted in the range $0.2 < m_T - m_{\pi} < 0.7$ GeV/c$^2$ using the formula $dn/(m_T dm_T) = C \cdot exp (-m_T / T)$. The resulting inverse slope parameters ($T$) are presented in Fig.\ref{T_BeBe} for three SPS energies. Here, it is worth to remind that in the case of a static source the inverse slope parameter ($T$) is equal to the thermal (kinetic) freeze-out temperature ($T_{fo}$). 
In the case of an expanding source $T$ is equal to the freeze-out temperature ($T_{fo}$) plus the effect of the radial flow. The NA61 results show that at middle SPS energy (40$A$ GeV) inverse slope parameters are similar for $p+p$ and $Be+Be$ collisions ($T_{Be+Be} \approx T_{p+p}$). In contrast, at the top SPS energy $T_{Be+Be}$ values are significantly larger than $T_{p+p}$. This result is treated as a possible evidence of the transverse collective flow in $Be+Be$ collisions at higher SPS energies (Beryllium looks heavy at the top SPS energy). The results from the analysis of other particles (kaons, protons, etc.) are expected soon and, in principle, complete Blast-Wave Model fits, as well as {\it kink}, {\it horn}, {\it step}, and {\it dale} plots should be available also for $Be+Be$ interactions.

\subsection{Looking for the critical point}

The critical point of strongly interacting matter can be nowadays treated as the Holy Grail of modern heavy-ion physics. We do not have solid conclusions from experimental results, and also theorists do not reach the agreement concerning the location and even the existence of the critical point (see review paper \cite{misha_CPpositions} and the recent lattice predictions above). Fluctuations and correlations can help to locate the critical point of strongly interacting matter. This is in analogy to critical opalescence (observed in most liquids, including water), where we expect enlarged fluctuations close to the CP. For strongly interacting matter a maximum of fluctuations is expected when the freeze-out (not the early stage!) happens near the CP. Therefore, the CP should be searched rather above the onset of deconfinement energy, found by NA49 to be $\sqrt{s_{NN}} \approx 7.6$~GeV \cite{na49pikp}.

Over the past years several experimental observables were proposed to look for the CP in heavy ion collisions. Among them are fluctuations of mean transverse momentum and multiplicity \cite{SRS}, pion pair (sigma mode) and proton intermittency \cite{inter_th}, etc. The NA49 experiment used the scaled variance of multiplicity distributions ($\omega$) and the $\Phi_{p_{T}}$ measure to quantify multiplicity and average $p_T$ fluctuations, respectively\footnote{In the case of no fluctuations/correlations $\Phi_{p_T}$=0, $\omega$=0. For a Poisson multiplicity distribution $\omega$ equals 1. The $\Phi_{p_T}$ is a {\it strongly} intensive measure of fluctuations (in thermodynamical models it does not depend on volume and volume fluctuations), whereas $\omega$ is an intensive one (it does not depend on volume but depends on volume fluctuations, and therefore NA49 used only 0-1\% most central data to study $\omega$ in $Pb+Pb$ collisions). See also \cite{kg_kielce2013} for other new {\it strongly} intensive measures of fluctuations used since recently by NA49 and NA61.} (see \cite{kg_qm09} for details).  
The position of the chemical freeze-out point in the ($T - \mu_B$) diagram can be varied by changing the energy and the size of the colliding system \cite{beccatini} as presented in Fig. \ref{phase_diagram} ($T_{chem}$ decreases from $p+p$ to $Pb+Pb$ interactions at the top SPS energy, and $\mu_B$ decreases with increasing energy of $Pb+Pb$ collisions). Therefore, NA49 analyzed the energy ($\mu_B$) dependence of $\omega$ and $\Phi_{p_T}$ for central $Pb+Pb$ collisions, and their system size ($T_{chem}$) dependence ($p+p$, central $C+C$, $Si+Si$, and $Pb+Pb$) 
at the highest SPS energy.



\begin{figure}[h]
\centering
\includegraphics[width=0.35\textwidth]{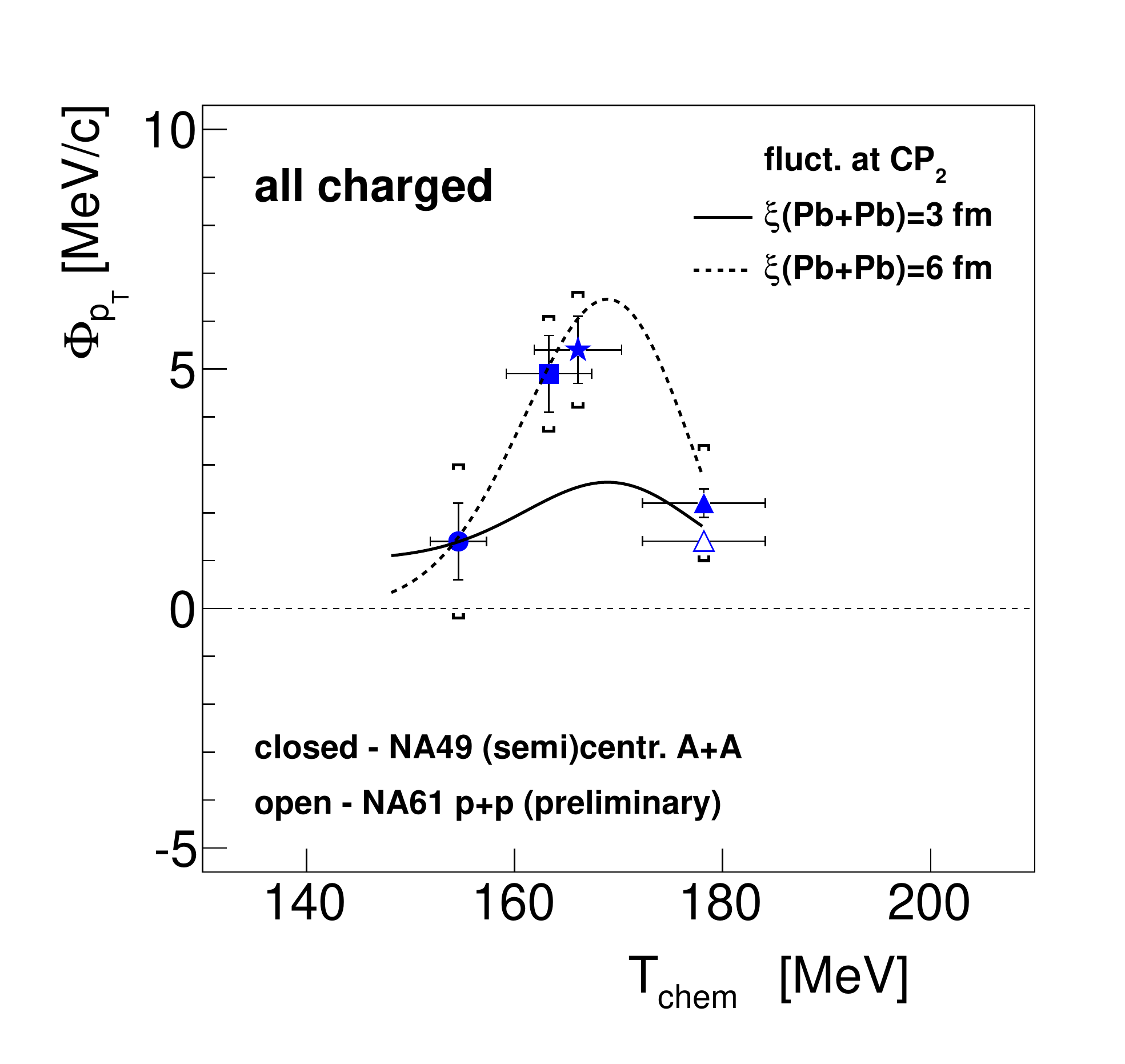}
\includegraphics[width=0.35\textwidth]{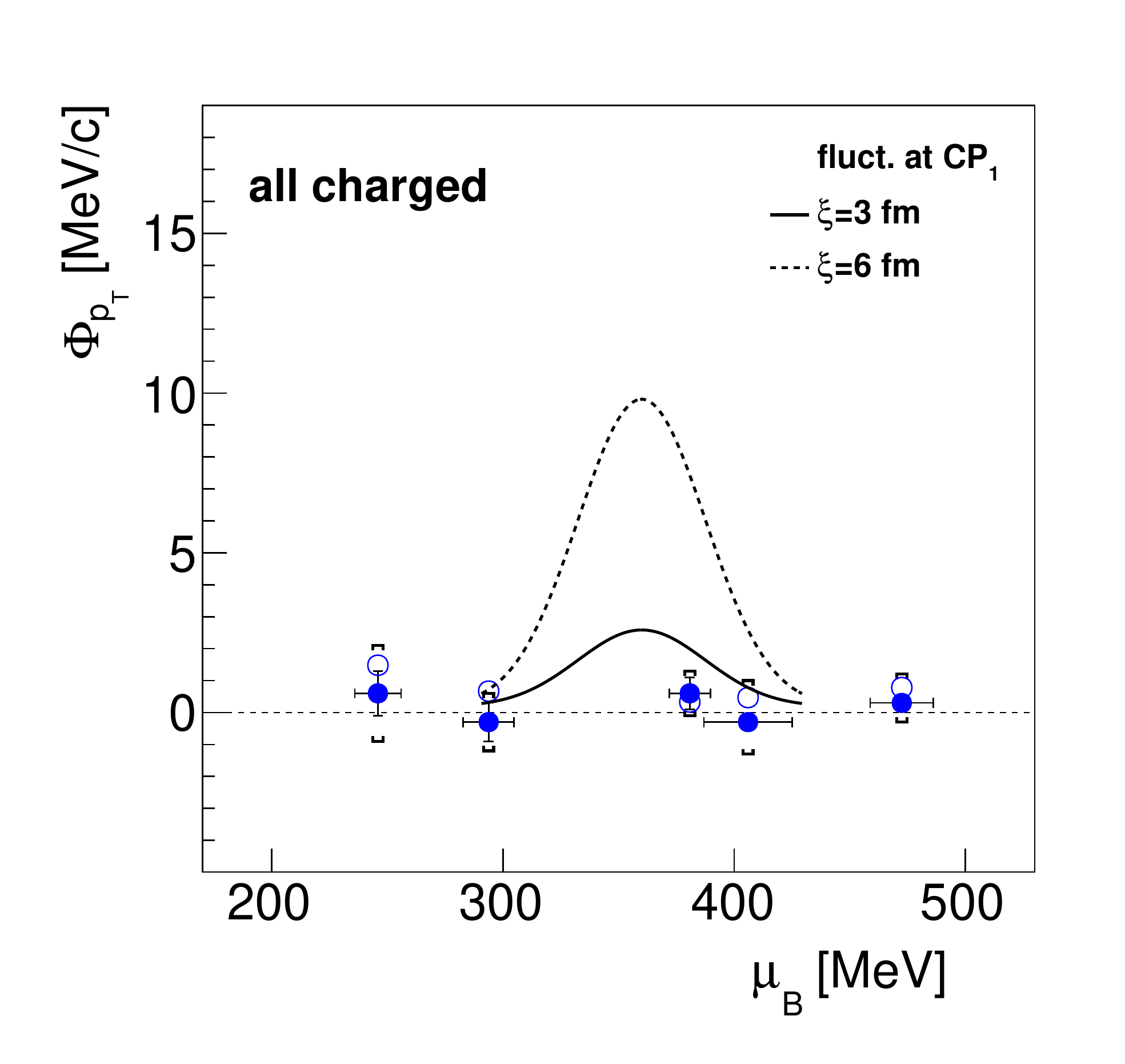}
\includegraphics[width=0.35\textwidth]{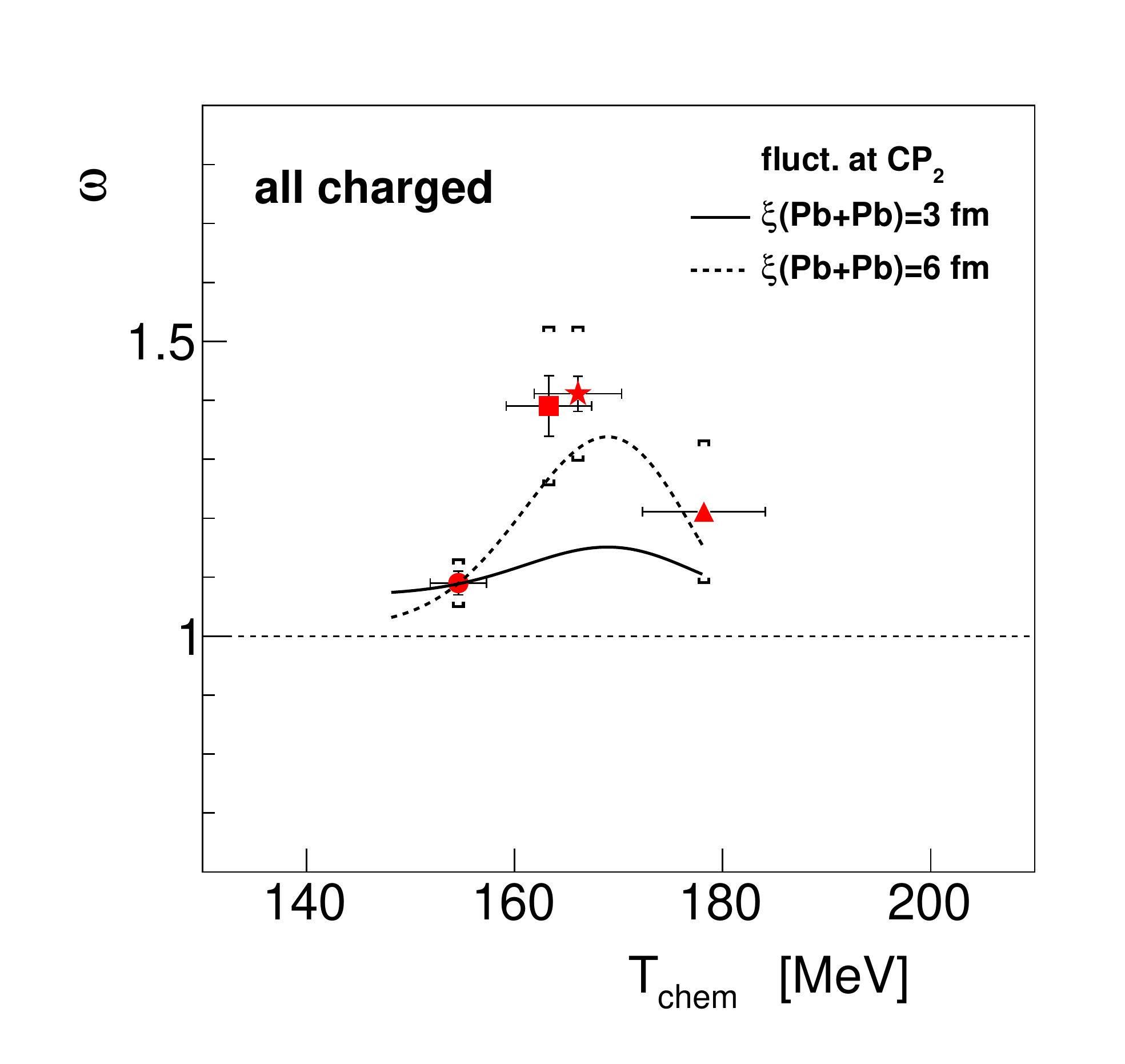}
\includegraphics[width=0.35\textwidth]{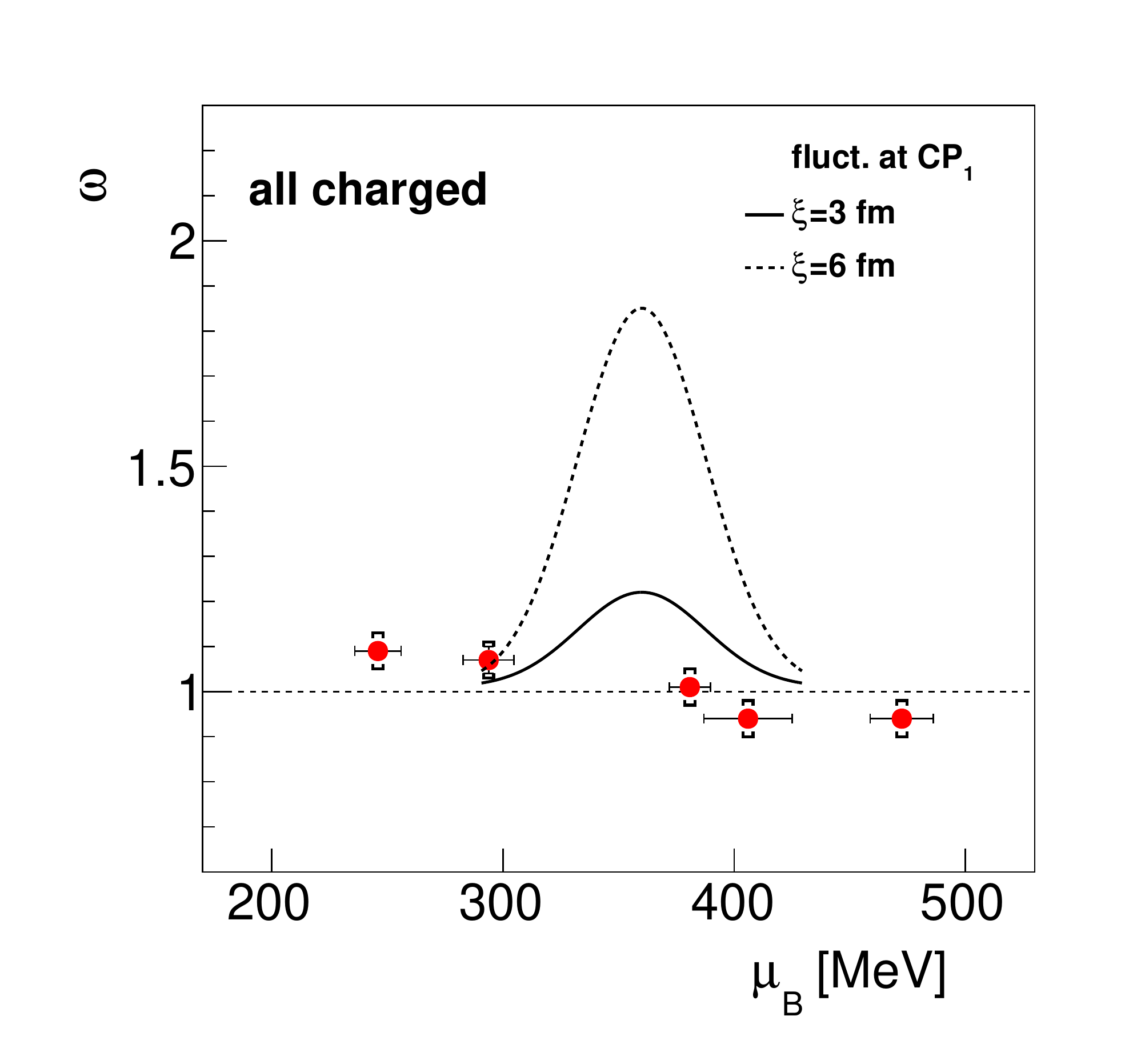}
\vspace{-0.3cm}
\caption[]{\footnotesize {Left panels: NA49 system size ($T_{chem}$) dependence ($p+p$, central $C+C$, $Si+Si$, and $Pb+Pb$) of transverse momentum ($\Phi_{p_T}$) and multiplicity ($\omega$) fluctuations at the highest SPS energy ($\sqrt{s_{NN}}=17.3$ GeV). Right panels: NA49 energy ($\mu_B$) dependence of $\Phi_{p_T}$ and $\omega$ for central $Pb+Pb$ collisions. NA61 preliminary $p+p$ results are shown as open points (for NA61 only statistical errors are shown). All values presented here (NA49 \cite{kg_qm09} and NA61 \cite{kg_kielce2013, PS_ISMD2013}) are obtained in the forward-rapidity region and in a limited azimuthal angle acceptance (for details see the corresponding papers in \cite{kg_qm09}).}}
\label{fipt_omega}
\end{figure}

Figure \ref{fipt_omega} shows that there are no indications of the CP in the energy dependence of multiplicity and mean $p_T$ fluctuations in central $Pb+Pb$ collisions (closed points, right panels). However, the system size dependence of both measures at 158$A$ GeV ($\sqrt{s_{NN}}=17.3$ GeV) shows a maximum for $C+C$ and $Si+Si$ interactions \cite{kg_qm09} (left panels, closed symbols). The peak is even two times higher for all charged than for negatively charged particles (not shown, see \cite{kg_qm09}) as expected for the CP \cite{SRS}. This result is consistent with a CP location near the freeze-out point of $p+p$ interactions at the top SPS energy ($T=$178 MeV, $\mu_B=250$ MeV; CP$_2$ in Fig. \ref{phase_diagram}). The theoretical magnitude of the CP effect has a maximum close to $Si+Si$ instead of $p+p$ system due to the fact that the correlation length in the model monotonically decreases with decreasing size of the colliding system (see \cite{kg_qm09} for details).

The critical point search will be continued by the NA61/SHINE experiment, where a two-dimensional scan of the phase diagram will be performed (see Fig. \ref{phase_diagram}). A maximum of CP signatures (fluctuations of multiplicity, average $p_T$, intermittency, etc.), so-called {\it hill of fluctuations}, is expected when the system freezes out close to CP. The first preliminary NA61 results from $p+p$ interactions \cite{kg_kielce2013, PS_ISMD2013} are shown in Fig. \ref{fipt_omega} as open points. There is no significant difference between the energy dependence of $p_T$ fluctuations for NA61 $p+p$ and NA49 central $Pb+Pb$ collisions (both experiments used the same NA49 phase-space cuts). There are no indications of critical point in the energy scan of $Pb+Pb$ and $p+p$ interactions and we are waiting for the results from $Be+Be$, $Ar+Ca$, and $Xe+La$.   

\begin{wrapfigure}{r}{8.cm}
\centering
\includegraphics[width=0.5\textwidth]{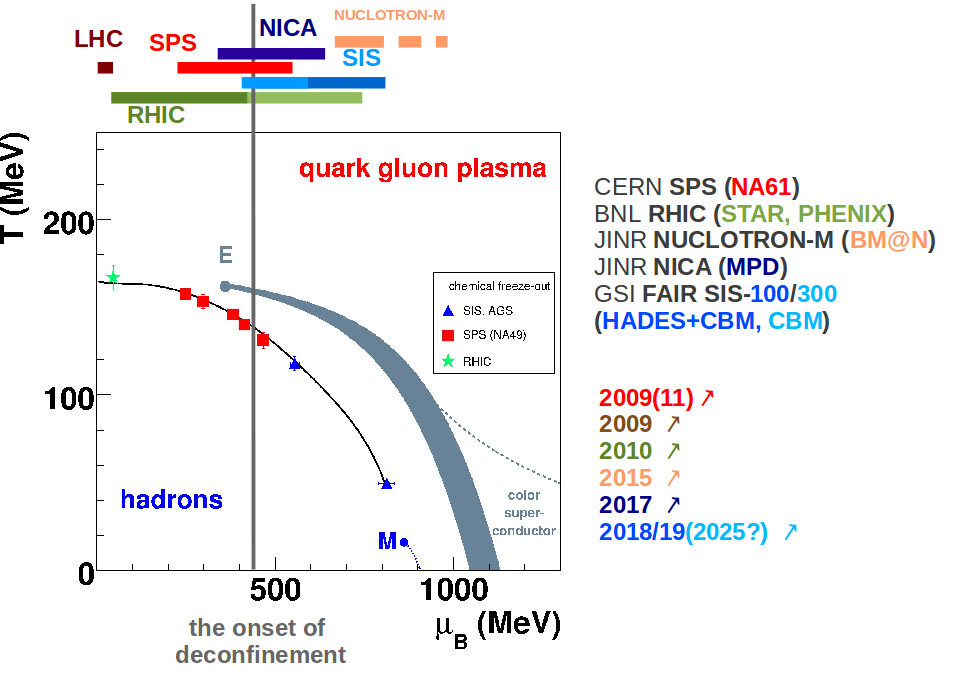}
\vspace{-0.2cm}
\caption[]{\footnotesize {The expected ranges of $\mu_B$ for existing and future heavy-ion experiments. Dark and light color for RHIC denote STAR plans for collider and fixed target modes, respectively. The highest RHIC energies (lowest $\mu_B$) have been already available since 2001/2002.}}
\label{future_exp}
\end{wrapfigure}

Other fluctuation measures were studied both by SPS and RHIC experiments, but the general conclusion is that the {\it energy} dependence of fluctuations for heavy-ion collisions ($Pb+Pb$/$Au+Au$) does not show any effects that can be attributed to CP. NA49 and NA61 did not find any possible signatures of CP when studying the energy dependence ($Pb+Pb$ and also $p+p$ collisions) of chemical (particle type) fluctuations \cite{cpod_kg, cpod_mmp, kg_sqm, anar_na49}. Also, electric charge \cite{delta_q} and azimuthal angle \cite{kg_sqm} fluctuations in central $Pb+Pb$ data did not show any unexpected behaviour. The RHIC BES studied: net-charge and net-proton fluctuations \cite{netch_netp_STAR}, $p_T$ correlations, particle ratio (chemical) fluctuations \cite{sahoo_wwnd2014}, but again, within the uncertainties, no clear non-monotonic behaviour, and thus no clear evidence of CP in the energy scan of $Au+Au$ collisions was observed. The $\mu_B$ ranges of the existing and future heavy-ion experiments are presented in Fig.~\ref{future_exp}.

\section{Summary}

At the {\it LHC and top RHIC energies we mainly study the properties of quark-gluon plasma}. The existence of QGP in $Au+Au$/$Pb+Pb$ collisions at high energies is well established, but there are some interesting measurements suggesting that the collectivity may be present also in high multiplicity $p+p$ and $p(d)+A$ collisions (but note that the collectivity does not necessarily mean QGP formation). Therefore, $p+p$ and $p(d)+A$ events should not be treated merely as a boring reference to $A+A$ interactions! The radial flow may be present also in light $Be+Be$ system at higher SPS energies.

At the {\it SPS and RHIC Beam Energy Scan energies we can study the transition region} between QGP and hadron gas. Several observables (RHIC BES and NA49: $dv_1/dy$, NCQ scaling of $v_2$, $R_{AA}$, {\it kink}, {\it horn}, {\it step}, {\it dale}, etc.) show that the energy threshold for deconfinement in $Pb+Pb$/$Au+Au$ collisions is located close to middle SPS energies. Is will be very interesting to check whether QGP may be created also in smaller systems (energy scan with small and intermediate mass systems in NA61). At {\it SPS and RHIC BES energies we can also look for the critical point of strongly interacting matter}. Fluctuations of average $p_T$, multiplicity, multiplicity of low mass $\pi^{+}\pi^{-}$ pairs and protons (see \cite{grecy} for the last two) tend to a maximum in $Si+Si$ collisions at the top SPS. It might be connected with CP at SPS energies, which is a strong motivation for future experiments. However, it is the beginning of the story! Much more effort is needed both from experimental (corrections, proper measures of fluctuations, etc.) and theoretical (lattice, models with predicted magnitudes of fluctuation measures at CP) side.

\vspace{1cm}

\noindent
{\bf Acknowledgements:} This work was supported by the the National Science Centre, Poland grant 2012/04/M/ST2/00816. Katarzyna Grebieszkow would like to thank the organizers 
of XXII. International Workshop on Deep-Inelastic Scattering and Related Subjects for inviting her and giving the opportunity to present the recent results from heavy-ion experiments.


\end{document}